\documentclass[11pt,a4paper]{article} 

\usepackage[utf8]{inputenc} 

\usepackage{geometry} 
\usepackage{graphicx} 

\usepackage{amsmath} 
\usepackage{amsthm}
\usepackage{amssymb}
\usepackage{epsfig}
\usepackage{color}
\usepackage{setspace}
\usepackage[font={small,it}]{caption}

\usepackage{booktabs}
\usepackage{multirow}
\usepackage[square,numbers,sort&compress]{natbib}
\usepackage{array} 
\usepackage{verbatim} 

\usepackage{color}

\newcommand{\ok}{\checkmark}
\newcommand{\ntick}{$\times$}
\newcommand{\tick}{\checkmark}

\title{
\textbf{
{\huge
Foreign Exchange Market Microstructure and the WM/Reuters 4pm Fix\bigskip
}
}
}

\date{05/02/2016}

\author{P. S. Michelberger\thanks{Corresponding author; contact e-mail address: PMichelberger@recordcm.com;}\,
 and J. H. Witte\vspace{.5cm}\\
Record Currency Management Limited \\ 
Morgan House, Madeira Walk \\
Windsor, Berkshire, SL4 1EP, UK \vspace{.5cm}\\
\bigskip
{\small To be published in: The Journal of Finance and Data Science}
}

\begin{document}

\maketitle

\begin{abstract}
A market fix serves as a benchmark for foreign exchange (FX) execution, and is employed by many institutional investors to establish an exact reference at which execution takes place. The currently most popular FX fix is the World Market Reuters (WM/R) 4pm fix. Execution at the WM/R 4pm fix is a service offered by FX brokers (normally banks), who deliver execution at the fix provided they obtain the trade order. 
In this paper, we study the market microstructure around 4pm. We demonstrate that market dynamics can be distinguished from other times during the day through increased volatility and size of movements. Our findings question the aggregate benefit to the client base of using the WM/R 4pm fix in its current form.

\vspace{0.5cm}

{\bf Keywords:} 
Finance, Financial Markets, Foreign Exchange Markets, Market Microstructure, 
Behavioural Finance, Exchange Rate Benchmarks, FX Execution, Market Manipulation,
WM/Reuters 4pm Fix, FX Market Microstructure
\end{abstract}

\section{Introduction}

Execution of foreign exchange transactions at a fixed market benchmark rate is common amongst institutional investors. Appeal lies in the fact that, particularly when tracking benchmark indices in other asset classes, currency conversion can be performed by the same underlying FX rates as employed by their target index. 
Furthermore, aggregation of fixing trades by market makers prior to execution increases liquidity, which, in theory, enables clients to trade large FX amounts at a trusted rate. 
Fixing trades therefore enable clients to trade without having to worry about best execution, such that they do not have to minimise the risk of moving the price when transacting in the market.
Moreover, as a market fix is uniform across providers, it is also intended to build trust in client/provider relationships, as a clearly defined and measurable service arises.

The currently most important benchmark rate is the World Market Reuters 4pm London fix 
(WM/R 4pm fix), 
which accounts for approximately $1$-$2\%$ of the total $\$2\,\text{trillion}$ daily volume in the FX spot market \cite{BIStriannual}.
 
The WM/R 4pm fixing rate \cite{WMR_fix} is determined using the methodology described in Section \ref{sec_WMRfix}. It is based on the last trade prices as well as the last best bid and offer quotes at the end of each one second interval between 15:59:30 to 16:00:30 GMT.
Therefore, to achieve the WM/R 4pm fix rate, service providers execute their fixing orders within this 60 second interval. 

Compression of large order flow into a narrow time window can be expected to give rise to a special market structure around the fixing time, which is part of what we will look at in this paper.

The described structure has recently come to the attention of the wider public, as concerns have been raised regarding market participants who may have used the construction mechanism of the WM/R 4pm fix to influence the benchmark \cite{US:CFTC,UK:FCA,Bloomberg_1}. 

Subsequently, multiple suggestions regarding how to improve the construction of the WM/R 4pm fix have been brought forward \cite{FSB_report, BoE_FEMR}.
These  investigations mainly focus on the mitigation of eventual manipulation possibilities. Generally, little research has been done on the observable market structure around the WM/R 4pm fix 
\cite{Evans:2014, NBER:2015}, 
despite the importance of the WM/R 4pm fix in the services received by many institutional investors.

In this regard, spot rate volatility and extreme spot rate movements are of particular interest, since they are important parameters in the quest for best execution practice. 
With best execution we refer to the trader's intention to execute as close as possible to the average market rate during the respective part of the trading day. 
High volatility and extreme spot rate movements increase the probability to trade at an outlier price. Such dynamics consequently introduce tracking error by pushing the realised transaction rates away from the sought after average rates. 

In previous work, Osler and Bruce (2008) \cite{Osler:2008}, Osler and Tanseli (2011) \cite{Osler:2011}, as well as Bjønnesa and Rime (2005) \cite{Bjonnesa:2005} have investigated the general interactions between clients and dealers, while the recent work by Chaboud et. al. (2013) \cite{Chaboud:2013} shifts the focus to the impact of algorithmic order execution.

In this paper, we investigate the influence of the WM/R 4pm fix on the dynamics of both spot rate volatility and extreme spot rate movements. Our study shows that the order compression around the fix indeed changes the market behaviour, resulting in spiking volatility within the WM/R 4pm fixing window, and an increased probability for spot rate extrema during this period.

\section{WM/R 4pm Fix Methodology\label{sec_WMRfix}}

Before discussing the market dynamics around the WM/R 4pm fix, a short description of its construction
shall be provided \cite{WMR_fix, FSB_report}. 
The WM/R 4pm fixing rate is determined for currency spot, forward, and non-deliverable forward rates. 
The calculation differs between forward and spot rates. While, for the former, a single rate snapshot at the fixing time is used as the benchmark, the spot rate calculation utilises several price quotes within an interval around the fixing time. 
In the following, we will focus on spot rates only.

Spot fixings are determined for 160 currency pairs, which are split into trade currencies for liquid pairs\footnote{
	Pairs containing only the following currencies: 
	AUD, CAD, CHF, CZK, DKK, EUR, GBP, HKD, HUF, ILS, JPY, MXN, NOK, NZD, PLN, RON, RUB, SEK, SGD, TRY, and ZAR.
}
and quote currencies for illiquid pairs.
For our analysis, we consider 12 trade currency pairs\footnote{
	We selected the pairs: 
	EURUSD, USDJPY, EURJPY, EURGBP, USDGBP, AUDUSD, EURCHF, USDCHF, GBPCHF, EURSEK, USDMXN, USDSGD. 
(see Section \ref{subsec_data}) which were selected to include the most liquid G10 currencies (USD, EUR, JPY). We furthermore selected two highly liquid EM pairs (MXN, SGD) and three less liquid G10 pairs (AUD, CHF, SEK), for which data was readily available.}.

To calculate the fix, WM/R sources data from Thomson Reuters, EBS, and Currenex.  
For the time period over which we investigate the WM/R 4pm fix, all trade currency data is obtained from a primary source\footnote{
	After chaning the methodology of the WM/R 4pm fix on the 15/2/2015, WM/R now sources currency data from multiple sources 
	and combines these datasets. 
}. 
For currency pairs involving CHF, EUR, and JPY, WM/R uses data from EBS. 
Currenex data supplements it as a secondary source, if the number of datapoints is too low. 
RUB data is solely obtained from EBS. 
Benchmark rates for other pairs are based on Thomson Reuters data only\footnote{
 	AUD, CAD, CZK, DKK, GBP, HKD, HUF, ILS, MXN, NOK, NZD, PLN, RON, SEK, SGD, TRY, and ZAR.
}.
The WM/R 4pm fix rate is obtained by accumulating quote and trade data within a set time interval around the fixing time.
For the time period prior to 15/02/2015, which we investigate here, this interval is 1-min. for trade currencies and 2-min. for quote currencies, i.e., for the WM/R 4pm fix 
the accumulation times are 15:59:30 - 16:00:30 and 15:59:00 - 16:01:00, respectively. 

Notably, following recent recommendations \cite{FSB_report}, WM/R implemented changes to its calculation methodology \cite{WMR_fix}, which became effective on 15/02/2015. 
After this date, the data sourcing window is widened to 5-min., 
 i.e., for the WM/R 4pm fix the accumulation time changes to 15:57:30 - 16:02:30. 
Additionally, trade currency data will be gathered from different sources. To this end, Thomson Reuters data is pooled together with EBS and Currenex data for pairs containing CHF, EUR, JPY, and RUB. 

Quote currency data is sampled at the end of 15-sec. intervals, leading to 9 datapoints in total. 
At each sampling point, the last valid best bid and ask quotes are recorded. Subsequently, the medians of these bid and ask quotes are calculated, whose mid-price yields the WM/R benchmark rate.

\begin{figure}[h!]
\includegraphics[width=\textwidth]{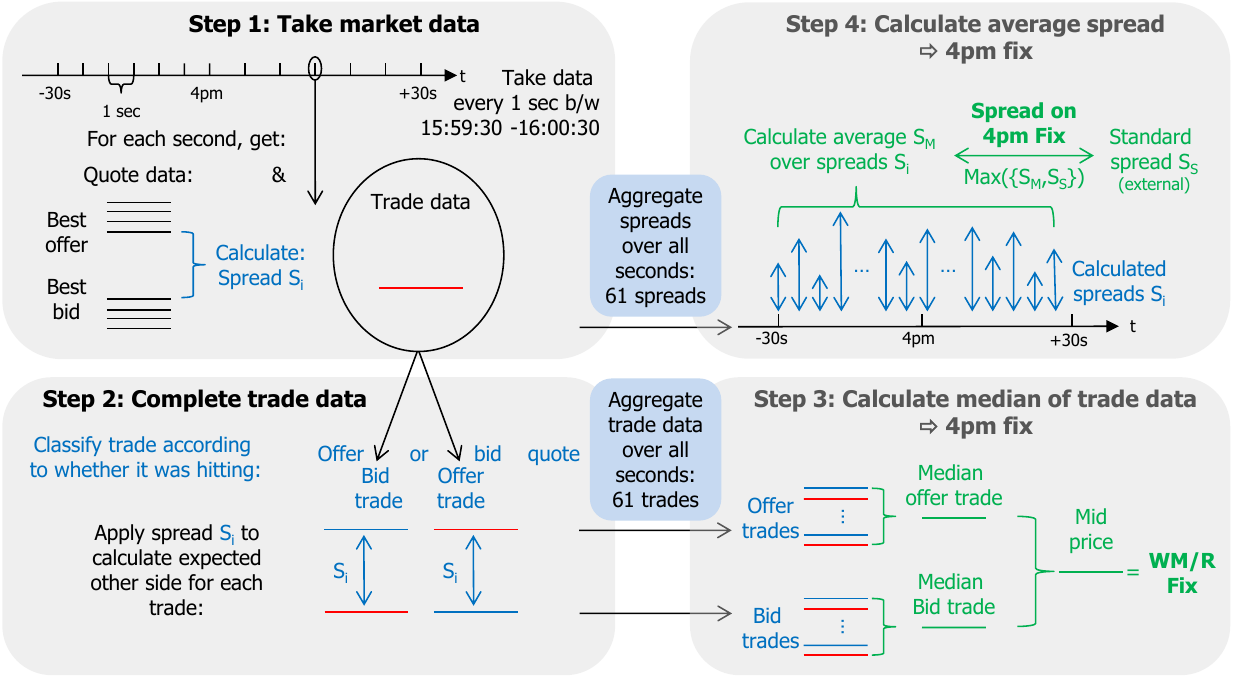}
\caption{Methodology for calculating the WM/R benchmark rate for trade currency spot rates.}
\label{fig_WMRmethod}
\end{figure}

For trade currencies, a more elaborate method is used, which is illustrated in Figure \ref{fig_WMRmethod}. 
Data is sourced at the end of each 1-sec. interval $i$, which amounts to 61 datapoints in total. 
The last valid best bid and ask quotes, as well as the last rate at which a trade happened in the interval are captured. 

To calculate the fixing rate for trade currencies, firstly, the spread $S_i$ of the best bid and offer quotes is determined for each interval. 
Subsequently, the trade rate is categorised into a bid trade or an offer trade, depending on whether the trade hit a quote on the bid or the offer side of the order book. 

The spread $S_i$ is applied to the trade rate to infer the opposite side of the order book, resulting in an inferred trade rate at that side. 
Trades that fall outside the best bid and offer quotes, i.e., trades which were executed hitting quotes inside the order book, are excluded from the calculation.
Combining the inferred and actual trade rates ideally yields subsets for the bid and the offer side containing a total of 61 datapoints each (bid trades and offer trades in Figure \ref{fig_WMRmethod}). 
Notably, this assumes that a valid trade occurs within each of the 1-sec. intervals, which is not necessarily the case. If there is insufficient trade rate data, bid and offer quotes are used instead. 
The exact limit on the number of required trade datapoints is discretionary to WM/R and unpublished.
In such a case, quote data is not pooled together from different data sources, but rather data from the source with the largest number of valid quotes is used\footnote{
In the event of an equal number of quote datapoints from two or more sources, a fix is calculated for each source individually and the final 
WM/R 4pm fix is obtained by taking the average over these individual fixing rates. Should there be only one quote datapoint available, which is obtained from two different sources, the source with the most recent quote update is used.
}.

For each side, i.e., bid trades and offer trades, the median of the rates is computed. The mid-price between the median bid and median offer trades yields the WM/R 4pm fix.  
Its spread is given by a predefined standard spread $S_S$ for each currency pair, which is supposed to reflect the liquidity at different times of the day. The determination of this standard spread is discretionary. However, if the market spread $S_M$, defined as the mean over all spreads $S_i$ obtained from the quote data, is larger than the standard spread, the market spread $S_M$ will be used instead. 
Applying this spread to the WM/R 4pm fix yields its bid and ask rates.

For both trade and quote currencies, the initially sourced data is subject to quality checks, which compare it to the general market level. This level is established from continuous market surveillance by capturing trade rates at every 15-sec. throughout the entire day. 
Quality checks happen automatically, testing for data consistency \cite{patent}, as well as on a discretionary basis, for which no detailed methodology has been published.

The mechanism for determining the WM/R 4pm fix closing spot rate suffers from two main weaknesses. On the one hand, it does not take into account volume information. 
Ignoring spot rate modifications arising from market impact, trade rates obtained from the execution of large orders contribute with the same weight to the benchmark as orders at the minimum size. On the other hand, capturing solely the last quote and trade rates at the end of a macroscopic time interval ignores market microdynamics. 
Market trends happening towards the end of each interval have the potential to affect the data used in benchmark calculation. 
For instance, the splitting of a large order into many transactions, distributed over the fixing interval, with each transaction selectively executed towards the end of the 1-sec. wide data aggregation intervals, can result in a modification of the benchmark in the direction of the order.


\section{Data Source and Available Data\label{subsec_data}}

For our analysis, we use spot rate data sourced from Bloomberg, for 12 currency pairs (see Section \ref{sec_WMRfix}). 
Due to data availability, the look-back periods vary between the currency pairs as follows:

\begin{center}
\begin{tabular}{l|l}
\toprule
01/2010-03/2014: &  EURUSD, EURJPY, EURGBP, EURCHF,\\
& GBPCHF, USDJPY, USDGBP, USDCHF\\
\midrule
01/2008-03/2014: & EURSEK, AUDUSD, USDMXN, USDSGD\\	
\bottomrule
\end{tabular}
\end{center}

All available data within these periods enters our analysis. 
However, we split the dataset into two subsets, the first running from the start date of each dataset until May 2013, and the second covering the period of June 2013 until April 2014. 
Our analysis is conducted separately on both sets of data. This is done because in June 2013 the WM/R 4pm fix attracted a lot of media attention \cite{Bloomberg_1}, which may have led to a change in the use of the fix by providers as well as clients. 
For this reason, we categorise the data into the time prior and post the media coverage of the WM/R 4pm fix.
The dataset for each currency pair contains at least three years of data before June 2013, at which point we anticipate a change in the market participants' execution behaviour.

All data has a minute-by-minute resolution, and contains four different price streams. These are the opening, the highest, the lowest, and the last spot rate\footnote{HLOC, high-low-open-closed.} obtained within each 1-min. interval. Since the opening rate corresponds to the closing rate of the previous point, our analysis only focuses on the latter three time series. 

Each interval is centred at the minute, i.e., datapoints are, for example, obtained at 15:59, 16:00 and 16:01.
The time indices are expressed in local London time\footnote{London time corresponds to GMT+(1-hour) during the period from the last Sunday in March to the last Sunday in October, and GMT otherwise.}. 

We investigate times from 01:01 to 22:59 every trading day. 
For the parts of your analysis that rely on the comparison between rates data at different minutes throughout the day 
(Section \ref{sec_extr}), we require a complete set of data entries during this period. 
This means that the daily dataset must have an entry for each minute within this period. 
Sporadically, daily datasets do not fulfill this criterion.
Datasets for days when this criterion is not met are excluded from this analysis.


\section{(In)consistency of Realised Volatility\label{sec_vol}}

As a first step in the examination of price dynamics, we investigate realised price volatility.

We observe the realised average volatility $\sigma$ at each minute within a day by first calculating the arithmetic minute-by-minute returns
$$
R(t)=\frac{\left(S(t)-S(t-1\,\text{min.}) \right)}{S(t-1\,\text{min.})}
$$
for the spot rate $S(t)$ in each 1-min. interval $t$. 
This is done for all minute datapoints contained in each day within the available datasets.

\begin{figure}[h!]
\centering
\includegraphics[width=0.8\textwidth]{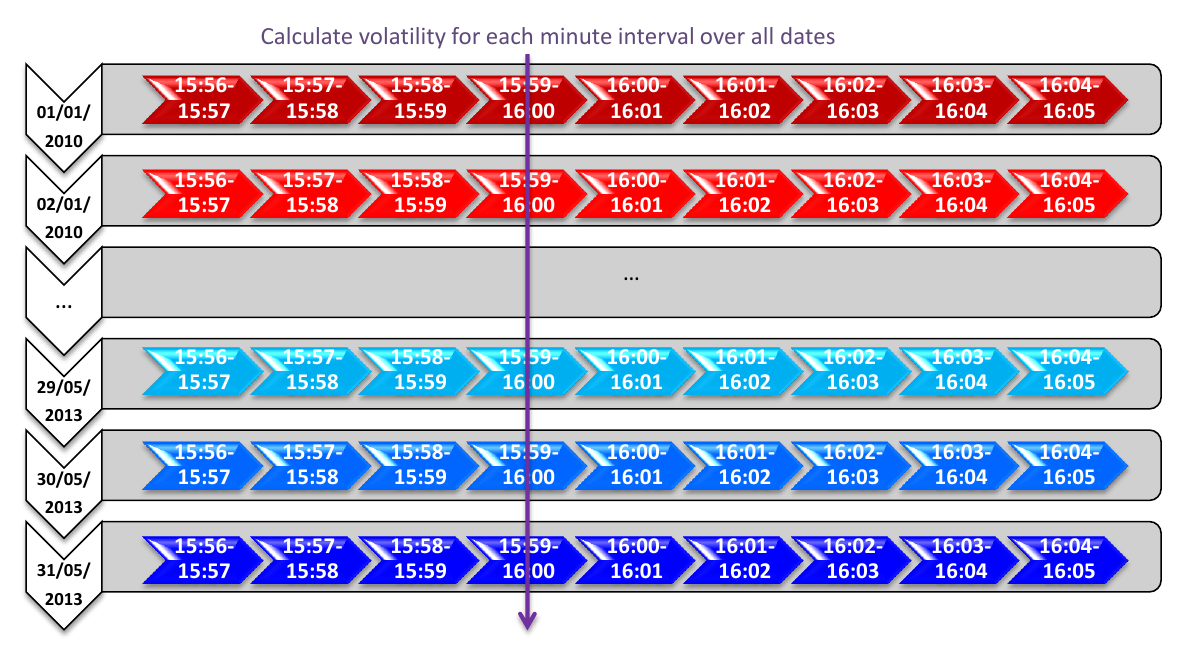}
\caption{
Method for calculating the average volatility $\sigma(t)$. Minute-by-minute returns are split-up into a matrix with the daily time index running along columns and the time series for each day in the dataset along rows.  
As indicated by the purple arrow, $\sigma(t)$ is obtained by determining the standard deviation over each column of the matrix.
}
\label{fig_vol_method}
\end{figure}

As shown in Figure \ref{fig_vol_method}, the time series of returns can be split-up as a matrix, where the time within each day is running along columns, and the dates contained in the respective dataset are running along rows.

We obtain average volatilities $\tilde{\sigma}(t)$ by, for each minute $t$,  calculating the standard deviations
of the returns $R(t)$ over all days.  Subsequently, we annualise\footnote{
	Assuming $252$ business days, the annualisation factor is $\alpha = \sqrt{252 \cdot 24 \cdot 60}$, yielding annualised volatilities $\sigma(t) = \alpha \cdot \tilde{\sigma}(t)$.
} these average volatilities and rename them to $\sigma(t)$.

\begin{figure}[h!]
\centering
\includegraphics[width=0.9\textwidth]{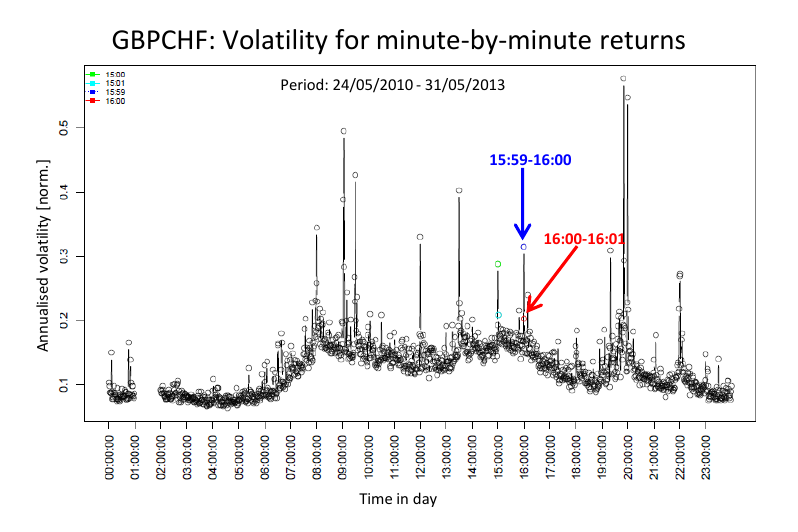}
\caption{
Average volatilities $\sigma(t)$ of minute-by-minute returns for the currency pair GBPCHF. Each minute is represented by one datapoint. The green point corresponds to 15:00-15:01, the cyan point to 15:01-15:02, the blue point to 15:59-16:00, and the red point to 16:00-16:01. The former pair is the volatility just after information release in the US at 10am EST. The latter pair are the minutes before and after the WM/R 4pm London fix.
}
\label{fig_vol_res}
\end{figure}

The resulting values for $\sigma(t)$ contain a rich structure of local extrema in volatility for single minutes in the day, standing out of a broad, smooth background distribution. 
Figure \ref{fig_vol_res} illustrates this by the example of the pair GBPCHF. The findings for all investigated currency pairs are summarised in Table \ref{tab_vol}.
The time locations for most spikes in volatility differ between the various currency pairs.
However, some local extrema occur consistently across all investigated pairs. 

Amongst the consistent extrema are the minutes 15:59-16:00 (blue point in Figure \ref{fig_vol_res}) and 16:00-16:01 (red point  in Figure \ref{fig_vol_res}), which are the minutes just before and after the WM/R 4pm fix. 
Here, particularly the minute leading up to the fix shows a significant increase in volatility compared to the minutes in the $\sim 50$-min. beforehand. 

A second set of consistently elevated volatility occurs at 15:00-15:01 and 15:01-15:02. 
This sudden volatility increase happens right after the release of market-relevant information \cite{Wikstroem14} 
in the US and the expiry of FX options at 10am EST.

It is important to note here that the increase in $\sigma(t)$, related to the WM/R 4pm fixing window, occurs in anticipation of the WM/R 4pm fix. 
This means the volatility spikes in the minute before 4pm. At this time, transaction orders for execution at the fix start to be processed. 
The volatility tails off just after 4pm, to an extent where, for some currency pairs, no pronounced local extremum exists for 16:00-16:01 (see Table \ref{tab_vol}). 

Changes in market dynamics consequently start before the event, and disappear directly afterwards.
By contrast, points indicating volatility related to market-relevant information arrival are often located after the event, once the information has been released.
We will discover a similar structure also for other metrics in Sections \ref{sec_extr} and \ref{sec_corr} below.
We emphasise that the presence of both sets of points is subset consistent, i.e., calculating the volatilities for parts of the data, spanning one year only, still allows for the observation of the same volatility spike structure.

Investigation of the second data category for the time period after May 2013 reveals a similar volatility structure.
These datasets also show a sudden change in market dynamics for the WM/R fixing window, returning back to the initial state thereafter.

\begin{table}[h!]
\begin{center}
  \begin{tabular}{ l | c | c| c | c | c | c }
    \toprule
Pair 	& 	\multicolumn{3}{ c |}{15:59-16:00} 			& 	\multicolumn{3}{ c }{16:00-16:01} \\
\midrule
	&	Start-2013 &	2013-2014 & Subset	&	Start-2013 &	2013-2014 & Subset \\	
\midrule
AUDUSD	&	\ok	&	\ok	& 	\ok	&	\ntick	& \ntick	& \ok 		\\
USDCHF	&	\ok	&	\ok	& 	\ok	&	\ntick	& \ntick	& \ok 		\\
USDGBP	&	\ok	&	\ok	& 	\ntick	&	\ok	& \ntick	& \ok 		\\
USDJPY	&	\ok	&	\ok	& 	\ok	&	\ntick	& \ntick	& \ok 		\\
USDMXN	&	\ntick	&	\ok	& 	\ntick	&	\ntick	& \ntick	& \ok 		\\
USDSGD	&	\ok	&	\ok	& 	\ok	&	\ok	& \ntick	& \ntick 	\\
EURCHF	&	\ok	&	\ok	& 	\ok	&	\ntick	& \ntick	& \ntick 	\\
EURGBP	&	\ok	&	\ok	& 	\ok	&	\ok	& \ok		& \ntick 	\\
EURJPY	&	\ok	&	\ok	& 	\ok	&	\ntick	& \ntick	& \ok 		\\
EURSEK	&	\ok	&	\ok	& 	\ok	&	\ok	& \ok		& \ok 		\\
EURUSD	&	\ok	&	\ok	& 	\ok	&	\ntick	& \ntick	& \ok 		\\
GBPCHF	&	\ok	&	\ok	& 	\ok	&	\ntick	& \ntick	& \ntick  	\\
\bottomrule		
  \end{tabular}
\end{center}
\caption{
Observation of spikes in volatility $\sigma(t)$ in the minutes 15:59-16:00 prior to, and 16:00-16:01 post the WM/R 4pm fix. The first two columns for each minute show the observations in both dataset categories, from 2008 or 2010 to 05/2013, and from 06/2013 to 04/2014. The last column denotes the yearly subset consistency for the 2008/2010-2013 dataset. 
Subset consistency refers to the presence of the effect in subsets of the entire dataset, where each such set spans the time period of one year.
}
\label{tab_vol}
\end{table}

\section{Spot Rate Movements around 4pm \label{sec_extr}}

In case of the WM/R 4pm fix, a large amount of trading volume is confined into a small time window, which results in a significant transaction frequency increase. 

From a localised trading activity extremum with equal liquidity supplied on both sides of the order book, i.e., at bid and ask, one would expect an elevated frequency of transactions which show small spreads around some mean price. Such dynamics would lead to a reduction in volatility.

However, from unequal liquidity supply, one can expect that the surplus of orders on one side depletes the order book on the opposite side, leading to large spot rate movements. 
Such extreme movements also increase the volatility due to their larger deviation from the time interval's mean spot rate. 

In this section, we investigate the type of spot rate movements that actually underlie the observed volatility increase. To this end, we study the distribution of extreme price movements within the average trading day in our sample.

We will show here that there is indeed a significant increase in the probability for spot rate extrema around 4pm, and that the movement sizes are, on average, larger than their comparable counterparts at other hours during the day.  
From our findings, we will furthermore be able to conclude that the timing of such events differs between their occurrence around the WM/R 4pm fix and other types of events (such as information arrival).

\subsection{Analysis\label{subsec_extr_analysis}}
Similar to the previous section, we analyse each currency pair separately by splitting up each dataset into daily subsets. We examine the time series obtained from the lowest, the highest, and the last quoted prices with a 1-min. resolution.

\begin{figure}
\centering
\includegraphics[width=0.8\textwidth]{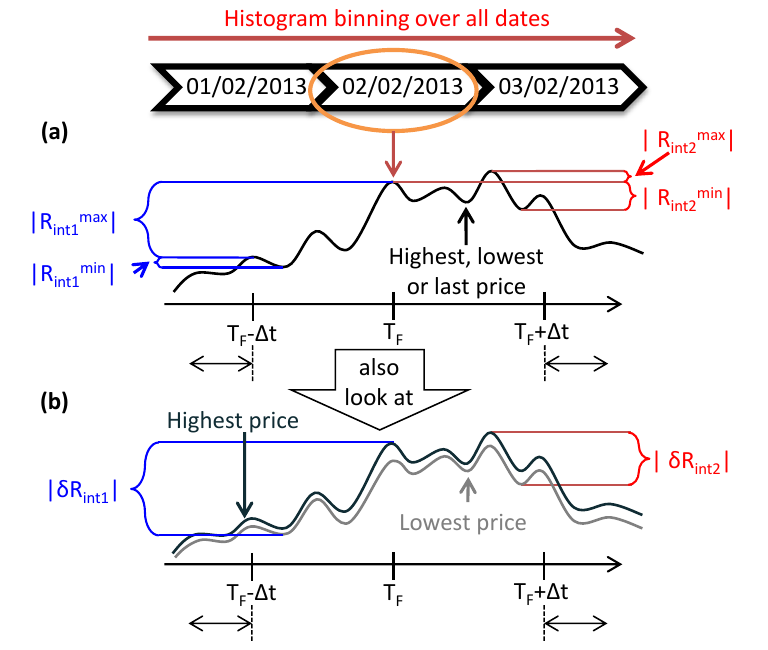}
\caption{Methodology for obtaining: (a) the maximum and minimum spot rate in intervals int1=$\left[ T_F - \Delta t, T_F \right]$ and int2=$\left[T_F , T_F+ \Delta t\right]$ around a fixing hour $T_F$, see main text for details; (b) 
the difference between the maximum of the highest spot price and the minimum of the lowest spot price within both intervals.}
\label{fig_2}
\end{figure}

Extrema in price movements are identified by observing the respective spot rates $S_d(t)$ in an interval $\Delta t$ before and after a fixed time $T_F$ for each day $d$ in the dataset, which contains a total of $N_d$ days. 
Here, $t$ denotes every 1-min. time step in the time series.
As Figure \ref{fig_2} (a) illustrates, the two intervals represent the time periods
\begin{align*}
\text{int1} =&\ \left[ T_F - \Delta t, T_F \right], \text{ and}\\
\text{int2}=&\ \left[ T_F, T_F + \Delta t \right].
\end{align*}
For each interval of size $\Delta t$, we initially evaluate the returns of the current sport rate ($S_d(t)$) with respect to the spot price at the start of each interval ($S(T_F-\Delta t)$) to obtain the \textit{period returns}
\begin{align*}
R_\text{int1}(t,T_F,\Delta t, d) =&\ \frac{S_d(t) - S_d(T_F-\Delta t)}{S_d(T_F-\Delta t)}, \text{ and}\\
R_\text{int2}(t, T_F, \Delta t, d) =& \frac{S_d(t) - S_d(T_F)}{S_d(T_F)}
\end{align*}
for intervals int1 and int2, respectively. 
We find the maximum and minimum of each set of period returns
\begin{align*}
R^\text{max}_\text{i} (T_F,\Delta t, d) &= \text{max} \left\{ R_\text{i} (t, T_F, \Delta t, d) \right\}, \text{ and} \\ 
R^\text{min}_\text{i} (T_F,\Delta t, d) &= \text{min} \left\{ R_\text{i} (t, T_F, \Delta t, d) \right\},\, 
\text{ for } 
i \in \left\{ \text{int1, int2} \right\},
\end{align*}
which represent the most extreme price movements inside each interval $\Delta t$ prior and post the time $T_F$. 

Additionally, we also determine the difference between both extrema 
\begin{equation*}
\delta R_i (T_F, \Delta t, d) = R_i^\text{max}(T_F, \Delta t, d) -R_i^\text{min}(T_F, \Delta t, d), 
\end{equation*}
where $R_i^\text{max}$  
is taken from the highest, and $R_i^\text{min}$ 
is taken from the lowest spot prices in each minute, as shown in Figure \ref{fig_2} (b).

Note that the time leading up to the fixing hour $T_F$ ($i = $ int1) and the time thereafter ($i = $ int2) are evaluated separately, i.e., we run the analysis for both intervals $i$. 

The size of these local extrema can be compared across different fixing times $T_F$ within the day, where we choose $T_F$ as each full hour ranging from 02:00 to 23:00.
To this end, we identify the time location of the most extreme movement $\tilde{R}^j_i(\tilde{T}_F,\Delta t,d)$ within each day, i.e., the hour $\tilde{T}_F$ 
for which the largest value $|R^j_i(T_F, \Delta t, d)|$ for all fixing hours $T_F$, can be observed: 
\begin{align*}
\tilde{R}^j_i(\tilde{T}_F,\Delta t,d) 		&= \underset{T_F}{\text{max}} \left\{ |  R^j_i ( T_F, \Delta t, d) | \right\}, 
\quad \text{and} \\
\delta \tilde{R}_i(\tilde{T}_F,\Delta t,d) 		&= \underset{T_F}{\text{max}} \left\{ \delta R_i ( T_F, \Delta t, d)  \right\}, 
\quad \text{ with } 
j \in \left\{ \text{max}, \text{min} \right\}. 
\end{align*}
So, for each day $d$, the hour $\tilde{T}_F$, at which the most extreme movement occurs, is determined. 
To illustrate the resulting set of variables and the typical sizes of interval returns, 
a comparison between the extreme interval returns $\tilde{R}_i^j(T_F,\Delta t,d)$ and the regular interval 
returns $R_i^j(t,T_F,\Delta t,d)$ is presented in the Supplementary Information \ref{app_size_ext_int_ret}. 

Knowing $\tilde{T}_F$, in a next step we establish the distribution of the occurrence frequency of these global daily extrema
$\tilde{R}^j_i(\tilde{T}_F,\Delta t,d)$  
against their hour of occurrence, $\tilde{T}_F$, over all $N_d$ trading days in the time series for the currency pair.

To this end, the times at which an hour $T_F$ corresponds to the location $\tilde{T}_F$ of the most extreme spot movement within the day are summed up for all days in the dataset. 
The histogram of these counts as a function of the fixing times $T_F$ yields the probability distribution for the occurrence of extreme price movements during each day. 
Mathematically, this probability $P(T_F,\Delta t)$ is given by:
\begin{equation}
P(T_F,\Delta t) = \frac{\underset{d \in N_d}{\sum}c(T_F,\Delta t, d)}{N_d}, 
\label{eq_ret_ext_prob}
\end{equation}
with the counts
\begin{equation*}
c(T_F,\Delta t,d) =
\begin{cases}
    1, & \text{ if } |\tilde{R}^j_i (T_F = \tilde{T}_F, \Delta t,d) | >  |\tilde{R}^j_i (T_F \neq \tilde{T}_F, \Delta t,d)|  \\
    0, & \text{ otherwise.}
\end{cases}
\end{equation*}

\noindent
Clearly, for a completely random market, such a histogram would be uniform. 
If the spot prices were to follow a random walk, the sizes of spot movements would, on average, not depend on the location along their paths where period returns are evaluated. 
Return extrema would consequently not cluster at any point in time, and the probability for extreme spot movements would be equally distributed between all hours $T_F$.

Therefore, any patterns in the histograms versus the hours $T_F$ indicate non-random market dynamics at specific times. Only such events can cause localised maxima in the probability distributions of extreme spot movements.

In a random market, the distribution should also not depend on either of the interval sides. 
Thus, the histograms are expected to show no differences between intervals int1 and int2. Neither would one expect a dependence on $\Delta t$.

The last point is of particular importance, as the interval size $\Delta t$ is a measure of locality of extreme movements. 
By decreasing $\Delta t$ from $59$-$\text{min.}$ down to $1$-$\text{min.}$, any extrema must be located successively closer to the fixing point $T_F$ to show up in the resulting histograms, as otherwise they would not be considered in the series of period returns $R^j_t(t, T_F, \Delta t, d)$. 

Consequently, if the market is not random and extreme events are particularly frequent just before (int1) or after (int2) a specific fixing time $T_F$, then one expects to observe a spike in the probability distribution at this time $T_F$, which disappears as the size of $\Delta t$ is increased.

The resulting histogram distributions are three dimensional, showing the probability $P(T_F,\Delta t)$ for extreme events over all hours $T_F$ and all interval sizes 
$1$-$\text{min.}\le \Delta t \le 59$-$\text{min}$.
The evaluation is conducted for all variables: 
$R^\text{max}$ for the highest and last prices, $R^\text{min}$ for the lowest and last prices, as well as for the maximal distance $\delta R$ between the highest and the lowest prices.

\subsection{Observations\label{sec_ext_ret_distr_observations}}

\begin{figure}
\centering
\includegraphics[width=\textwidth]{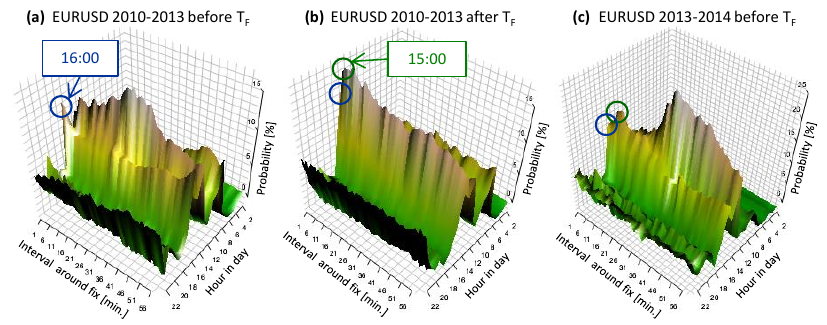}
\caption{Probability distributions $P(T_F, \Delta t)$ of extreme spot period returns as a function of fixing hour $T_F$ and interval size $\Delta t$ for EURUSD. 
(a) Extreme period returns in interval int1=$\left[T_F-\Delta t, T_F \right]$ for data from 05/2010 to 05/2013. (b) Extreme period returns in interval int2=$\left[T_F, T_F +\Delta t\right]$ for data from the same time period as in (a). 
(c) Extreme period returns in interval int1=$\left[T_F-\Delta t, T_F \right]$ for data from 06/2013 to 04/2014.}
\label{fig_3}
\end{figure}

Figure \ref{fig_3} displays an example for the obtained probability distributions for the currency pair EURUSD against the interval size $\Delta t$ and each hour $T_F$ in the day.
Every point on the surfaces in Figure \ref{fig_3} equals the probability for the most extreme intraday spot movement to happen in an interval of size $\Delta t$ before (Figure \ref{fig_3} (a) and (c)) or after (Figure \ref{fig_3} (b)) each hour $T_F$.  

Contrary to a random market, the distributions clearly show a pronounced structure. 
Most importantly, Figure \ref{fig_3} (a) shows a sharp increase in the $P(\Delta t,T_F)$, and therewith in the number of extreme events, just before $T_F = \text{16:00}$, which is observed for small interval sizes on the order of $\Delta t \sim 1\text{-min}$. It completely disappears and merges into an approximately constant background for larger intervals $\Delta t$. The exact timing varies between currency pairs, but it is generally between $1\text{-min.} \le \Delta t \le 5\text{-min}$. 

The data for the other $11$ pairs studied in this article show similar probability distributions to the EURUSD example in Figure \ref{fig_3}. These datasets are presented in Figures \ref{supp_fig_1}-\ref{supp_fig_3} of the Supplementary Information 
(see Section \ref{app_prob_ext_ret}). 

The sharp probability increase for narrowing window sizes $\Delta t$ suggests that, going into the quote accumulation window for the WM/R 4pm fix, extreme movements in the spot rate become significantly more likely than
during the time beforehand. Comparing the size of the probability increase with those at a similar interval size at other times $T_F$  in the day illustrates the extent of this difference\footnote{
	Note: 14:00 GMT, which also shows a probability maximum in Figure \ref{fig_3} (a), corresponds to 9am EST. This represents the onset of trading activity in the US and the aftermath of US economic report releases.
}.

Spot price movements before 4pm London time have, therefore, an increased likelihood to be global extrema.

The 4pm probability spike is not globally maximal in all of the investigated currency pairs. However, it is at least a local probability extremum for nearly all currency pairs (see Table \ref{tab_spikes} and Figures \ref{supp_fig_1} - \ref{supp_fig_3} in the Supplementary Information).
The onset of elevated trading activity for execution at the WM/R 4pm fix thus drives the spot rate such that demand on one side penetrates deeper into the order book on the other side.

For most currency pairs, these features are absent in the interval int2, 
which is looking at spot movements after the fixing times $T_F$, i.e., for the period $\left[T_F,T_F+\Delta t\right]$  (see Figures \ref{supp_fig_1} - \ref{supp_fig_3} in the Supplementary Information).
Illustrated in Figure \ref{fig_3} (b), we can see that the dominant contribution comes solely from $T_F=$15:00:00. 
Importantly, the first $30$-$\text{sec.}$ of int2 still fall into the $60$-$\text{sec.}$ WM/R quote accumulation window, 
which, for trade currencies,  
is given by the interval $T_F \pm 30$-$\text{sec.}$ (see Section \ref{sec_WMRfix}). 
Any trades happening within the first $30$-$\text{sec.}$ of the minutes after $T_F=$16:00:00 still contribute to the WM/R 4pm fix calculation. 
For this reason, it is sensible to assume that order flow and provision of liquidity are symmetric around $T_F$. 
The absence of probability spikes after $T_F$ does not necessarily mean that order book depletion stops abruptly at 16:00:00.
Instead, the interval return size can also be offset by the movements happening after 15:00 GMT, i.e. 
 $R^j_\text{int2}(T_F =\text{15:00},\Delta t = 1\,\text{min.},d) \ge R^j_\text{int2}(T_F =\text{16:00},\Delta t = 1\,\text{min.},d)$, 
where 15:00 GMT marks the FX option cut in New York and the release of US macroeconomic information at 10:00 EST in the US \cite{NBER:2015, Wikstroem14}.

Figure \ref{fig_3} illustrates another difference between spot rate movements induced by information arrival and that of trading at the benchmark.
Like the volatility extrema, observed in Section \ref{sec_vol}, the probability for extreme price movements around $T_F$=15:00 only increases after $T_F$, while it is low beforehand. 
Moreover, once information has been revealed, the induced increases in spot movement sizes are retained for longer time periods. The probability maxima in Figure \ref{fig_3} thus form ridges for $T_F=$14:00 and $T_F=$15:00.
Particularly, they do not decay quickly with $\Delta t$ like the probability spikes at $T_F=$16:00. 

While almost all pairs show probability extrema before $T_F=$16:00 for the time period before $06/2013$,
in the recent time period from 06/2013-03/2014 
the feature has become less pronounced for some of the currency pairs. For instance, the probability for extreme returns prior to the WM/R 4pm fix experiences a significant reduction for the EURUSD pair, shown in Figure\ref{fig_3} (c). 

This raises two questions: 
First, does the dataset 2013/2014 allow for sufficient statistics to observe the effect? 
And second, is the feature just present in a subset of the 2008/2013 data, which happens to dominate the resulting distribution? 

To answer these questions, we have split the 2008/2013 time series into chunks of yearly data. 
By running the same analysis on each subset, we observe similar patterns with probability spikes at $T_F=$16:00. 
Moreover, almost all currency pairs show subset consistency, i.e., probability extrema are observable in each yearly subset (see Table \ref{tab_spikes}).

We can thus conclude that the trading activity around the WM/R 4pm fixing window has, over the past, led to extreme movements for the spot rates. 
These extrema are localised in a short, minute-sized interval around the fixing time at 4pm. Their localisation sets them apart from other market effects, such as information release, which lead to a probability increase distributed across longer time intervals, forming ``probability ridges".

\begin{table}[h!]
\begin{center}
  \begin{tabular}{ l | c  | c | c }
    \hline
\toprule
Currency 	& Spike at 4pm 		& 	Spike at 4pm  	& Subset consistent \\
pair		& 2008/2010-2013 		&	2013-2014	 	& 2008/2010-2013 \\ 
\midrule
AUDUSD 	& \tick				&	\tick			& \ntick			\\ %
USDCHF 	& \tick				&	\ntick			& \tick			\\%
USDGBP 	& \tick				&	\tick			& \tick			\\%
USDJPY 	& \tick				&	\tick			& \tick			\\%
USDMXN 	& \tick				&	\tick			& \tick			\\%
USDSGD 	& \tick				&	\tick			& \tick			\\%
EURCHF 	& \tick				&	\tick			& \tick			\\%
EURGBP 	& \tick				&	\ntick			& \tick			\\%
EURJPY 	& \tick				&	\tick			& \tick			\\%
EURSEK 	& \ntick			&	\tick			& \ntick			\\
EURUSD 	& \tick				&	\ntick			& \ntick			\\%
GBPCHF 	& \tick				&	\ntick			& \tick			\\ %
\bottomrule
 \end{tabular}
\end{center}
\caption{Observation of probability maxima (local or global) for the occurrence of extreme spot rate movements in the interval $\text{int1} = \left[ T_F - \Delta t, T_F \right]$ prior to the WM/R fixing time $T_F = \text{16:00:00}$.
The first and second column distinguish between the two data categories, prior and post 06/2013. The last column denotes the yearly subset consistency for observations prior to 06/2013.}
\label{tab_spikes}
\end{table}

\section{Size of Extreme Spot Rate Movements at 4pm \label{sec_corr}}

Having observed increased spot movement, we now consider the movement size.
To this end, we slightly modify the previously used method to allow for measurement of the average movement size. 
The modified approach will, additionally, allow us to pinpoint the occurrence of increased spot rate movements to exactly 4pm.

\subsection{Method\label{subsec_corr_method}}

We use a blend between our previous analysis and one of the metrics introduced in reference \cite{Wikstroem14}.
Again, period returns
\begin{equation*}
R_\text{int}(t) = \frac{S(t) - S(T_F - \Delta t)}{S(T_F - \Delta t)}
\end{equation*}
are evaluated for an interval of size $\Delta t$. However, this time, the interval size is fixed to $\Delta t = 20$-$\text{min.}$ and centred around a fixing time $T_F$, i.e., the spot rate is analysed in a time period $t \in \left[ T_F - \Delta t, T_F + \Delta t\right]$. 
Note that the intervals before and after $T_F$ are considered simultaneously, so no distinction is made between the time leading up to the fix and times thereafter.

\begin{figure}[h!]
\centering
\includegraphics[width=0.9\textwidth]{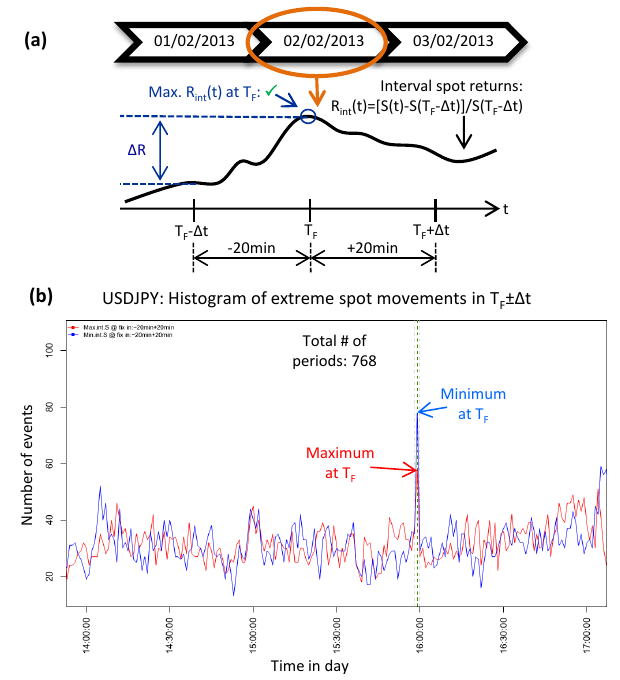}
\caption{
(a) Methodology for determining extreme interval returns $R_\text{int}(t)$ at the centre $T_F$ of an interval $\left[ T_F - \Delta t, T_F + \Delta t \right]$ of fixed size $\Delta t = 20\,\text{min}$.
The interval is moved consecutively over all minutes within each day. If the extremum in $R_\text{int}(t)$ is located at $\tilde{t} = T_F$, the movement size $\Delta R(\tilde{t})$ is counted. Histogram binning is performed over all days in the dataset, whereby the average over all resulting $\Delta R(\tilde{t})$ yields the extrema size.
 (b) Resulting histogram of extreme interval returns $\Delta R(\tilde{t})$ for the currency pair USDJPY. A clear increase in the occurrence of maxima (red line) and minima (blue line) is observed at $T_F = \text{16:00}$. The three vertical lines mark the times 15:59, 16:00 and 16:01, located around the WM/R fixing window. The underlying dataset contains daily spot rate data from 05/2010-06/2013.}
\label{fig_4}
\end{figure}

Within this interval of 40-min. total duration, we evaluate whether the maximum or minimum period return occurs exactly at time $t=T_F$, i.e. at the interval centre. 
This is done for $T_F$ set to all minutes within the day. So, contrary to the previous method, $T_F$
is moved through every point within the datasets for each day, not just through each full hour. 
Figure \ref{fig_4} (a) depicts this procedure.

For the resulting extrema, we firstly determine the size of the period return
\begin{equation*}
\Delta R(\tilde{t}) = |R^{j}_\text{int}(t = T_F)|,
\end{equation*}
whereby $\tilde{t}=T_F$ is the time index for the respective interval centre and $j \in \left\{\text{max},\text{min} \right\}$ denotes, whether the extremum at $\tilde{t}$ is the maximal or the minimal period return in the $40$-min. interval.
Given the fixed size for $\Delta t$, each of these extrema $|R^{j}_\text{int}(t = T_F)|$ happens exactly 20-min. after the interval starting point. For this reason, the sizes $\Delta R$ are comparable across all values of $t$.
On average, the sizes of these extrema amount to\footnote{
	This number represent the average over all minutes in the day and all $12$ currency pairs.
} 
$|\Delta R(\tilde{t})| \approx 5$ bpts., whereas the complete period return over the full 40-min. interval, i.e. $R(\tilde{t}+ \Delta t) = \frac{S(\tilde{t} + \Delta t) - S(\tilde{T} - \Delta t)}{S(\tilde{T} - \Delta t)}$, has an average size of $|R(\tilde{t}+ \Delta t)| \approx 3$ btps. 
However, as we will see in a moment, these sizes are larger for $\tilde{t}=$16:00, where we will observe 
$|\Delta R(\tilde{t} = \text{16:00})| \sim 10\,\text{bpts.}$.

By performing this analysis for each day within the dataset, we count the overall number of maxima and minima occurring at each minute $\tilde{t} = T_F$. 
Analogue to the methodology used in Section \ref{sec_extr}, we create histograms, showing the number of maxima/minima that have been registered at $\tilde{t} = T_F$ as a function of time $t$.

\subsection{Results\label{subsec_size_results}}

We discuss the results by means of example, considering the currency pair USDJPY. 
Figure \ref{fig_4} (b) shows the obtained histogram for $\Delta R(\tilde{t})$, with its time range focussing onto the afternoon hours around 4pm.
The analysis of the other 11 currency pairs yields similar results which are presented in the Supplemenatary Information (see Section \ref{app_hist_ext_ret_at_4pm}).

Comparison of the histogram counts between each minute illustrates the likelihood of extreme movements that occur at each specific minute $\tilde{t}$ when it is regarded as the fixing time $T_F$.
As expected from the results in Section \ref{sec_extr}, a sharp spike in the number of extreme events appears when the interval centre $T_F$ is at 16:00.
Here, the probability $P(\tilde{t})$ for the occurrence of the largest spot movement to happen at $\tilde{t} = T_F$ is increased to approximately $P(\tilde{t} = \text{16:00}) \approx15\,\%$, while the base level for all other times is approximately at $P(\tilde{t} \neq \text{16:00}) \approx 9.5\,\%$. 
For all other minutes, no pronounced structure stands out, so the likelihood of extrema can be assumed to be distributed randomly between these times.
The probabiliy for the occurrence of extreme interval returns $\Delta R(\tilde{t})$ is given by 
$$
P(\tilde{t}) = \frac{\sum_{j} N^j(\tilde{t})}{N_d} = \frac{N(\tilde{t})}{N_d},
$$ 
with $N^j(\tilde{t})$ denoting the total number of maxima and minima ($j \in \left\{\text{max},\text{min} \right\}$) in $\Delta R(\tilde{t})$ observed at time $\tilde{t} = T_F$, i.e. at the interval centre. 
$N_d$ is the total number of days in the sample. 
As Figure \ref{fig_4} (b) shows for the USDJPY example, these numbers are 
$N(\tilde{t}=\text{16:00}) \approx 115$, 
$N(\tilde{t}\neq \text{16:00}) \approx 73$, and 
$N_d = 768$, where $N(\tilde{t}\neq \text{16:00})$ is the mean number of extrema over all times $\tilde{t} \neq \text{16:00}$. 
Table \ref{tab_ExteremaNumber} lists these numbers for all of our 12 currency pairs alongside with the number of days in the sample and the resulting probabilities for extrema at the WM/R 4pm fix and other times in the day. 
Note that all listed values for $N(\tilde{t})$ represent the mean of the observed number of interval extrema for the last, highest and lowest price. 
The error on the number results from the standard deviation of these individual numbers around the quoted mean. 
The data for the individual price streams is listed in Section \ref{app_hist_ext_ret_at_4pm} of the Supplementary Information.

\begin{table}[h!]
\begin{center}
  \begin{tabular}{ l | c | c | c | c | c | c  }
    \toprule
 	&	\multicolumn{3}{c|}{Time $\tilde{t} = \text{16:00}$}		& 	 \multicolumn{3}{c}{Time $\tilde{t} \neq \text{16:00}$} \\
\midrule
Pair 	& 	$N(\tilde{t})$ &  $N_d$ & $P(\tilde{t})$ [\%] 	& $N(\tilde{t})$ &  $N_d$ 	& $P(\tilde{t})$ [\%]\\
\midrule
AUDUSD	 & $146 \pm 8$	& $1381$	& $10.6 \pm 0.6$	& $86 \pm 8$	& $1383 \pm 1$		& $ 6.2 \pm 0.5$	\\
USDCHF	 & $83 \pm 5$	& $766$	& $10.8 \pm 0.6$	& $54 \pm 5$	& $766.0 \pm 0.1$		& $7.0 \pm 0.7$	\\
USDGBP	 & $109 \pm 3$	& $777$	& $14.0 \pm 0.4$	& $84 \pm 10$	& $774 \pm 13$		& $10.8 \pm 1.2$	\\
USDJPY	 & $115 \pm 11$	& $768$	& $15.0 \pm 1.4$	& $73 \pm7 $	& $768.00 \pm 0.04$	& $9.5 \pm 1.0$	\\
USDMXN	 & $72 \pm 9$	& $1376$	& $5.2 \pm 0.6$	& $54 \pm 5$	& $1312 \pm 79$		& $4.1 \pm 0.5$	\\
USDSGD	 & $192 \pm 14$	& $1382$	& $13.9 \pm 1.0$	& $130 \pm 13$	& $ 1376 \pm 10$		& $9.4 \pm 0.9$	\\
EURCHF	 & $60 \pm 12$	& $766$	& $7.9 \pm 1.6$	& $36 \pm 4$	& $766.00 \pm 0.05$	& $4.7 \pm 0.5$	\\
EURGBP	 & $72 \pm 11$	& $766$	& $9.4 \pm 1.5$	& $31 \pm 4$	& $765.9 \pm 0.3$		& $4.0 \pm 0.5$	\\
EURJPY	 & $70 \pm 5$	& $743$	& $9.4 \pm 0.7$	& $42 \pm 5$	& $743$			& $5.6 \pm 0.6$	\\
EURSEK	 & $187 \pm 8$	& $1384$	& $13.5 \pm 0.5$	& $53 \pm 5$	& $1384 \pm 1$		& $3.8 \pm 0.4$	\\
EURUSD	 & $51 \pm 4$	& $702$	& $7.2 \pm 0.5$	& $40 \pm 5$	& $702.0\pm 0.3$		& $5.7 \pm 0.6$	\\
GBPCHF	 & $86 \pm 12$	& $738$	& $11.6 \pm 1.6$	& $38 \pm 4$	& $741 \pm 1$		& $5.2 \pm 0.6$	\\
\bottomrule
  \end{tabular}
\end{center}
\caption{Number of observed extreme interval returns $N(\tilde{t})$ in a 40-min. interval around a central time $\tilde{t}$, alongside their probability of occurrence $P(\tilde{t})$ in a sample containing $N_d$ days. 
The time $\tilde{t} = \text{16:00}$ denotes the WM/R 4pm fixing time, whereas the time $\tilde{t} \neq \text{16:00}$ represents the average of the observations over all other minutes within the trading day considered as the interval's centre. 
The presented data covers period 1, i.e. dates from 2008/10 to 2013, whereby we drop the restriction that each day must contain all minute entries.  Instead, we only require a complete datasets within the interval $\left[ \tilde{t} - \Delta t, \tilde{t} + \Delta t \right]$. 
See the Supplementary Information, Section \ref{app_hist_ext_ret_at_4pm}, for details.}
\label{tab_ExteremaNumber}
\end{table}

While this observation ties in with our previous findings, the present result is nevertheless different to the probability distributions presented in Figure \ref{fig_3}. 
Earlier, for a fixed $\Delta t$, the sizes of the spot movements were compared across all hours in a day, i.e., the period returns had to be a daily global extremum. 
Moreover, earlier, the extreme movements could happen anywhere within an interval, whereby the interval position was always set either prior or post $T_F$. 
Probability spikes therefore showed which time ranges contained the most dominant daily spot movements.
 
In the current analysis, the spot rate movements only have to be local extrema. But, they must be positioned exactly at the centre of a symmetric interval. 
The locality and symmetry requirements suppress effects from delocalised spot rate movements, such as those induced by information releases. 
We have seen in Figure \ref{fig_3} that these result in ``probability ridges'' instead of spikes, and therefore have an uncertainty regarding their timing. Since they also only occur after the information release event, there is no inherent symmetry with respect to any timing point $T_F$. 

Combining the results from both methods enables us to concisely distinguish between effects from participants' behaviour around a fixing time and other market effects. 

As we would expect from Table \ref{tab_spikes}, the probability spikes are present in the data for almost all investigated currency pairs.
Table \ref{tab_ExteremaSize} lists their average size $\Delta R(\tilde{t} = \text{16:00})$ obtained for maxima and minima. 
These are on the order of $10$ bpts., whereby the variation between currency pairs does not show any preference for more liquid or illiquid pairs.   

For clarification, these numbers are the average over all three investigated price quotes: the highest, the lowest, and the last price for each one minute interval. The individual values for all three types are identical to the mean within the error range, which derives from Gaussian error propagation on the individual errors of the three price types. 
The individual errors, in turn, are the standard error of the mean over all maxima or minima in $\Delta R(\tilde{t}=\text{16:00})$ at 4pm.

The data in Figure \ref{fig_4} (b) shows nearly parity between the number of extreme maxima and minima at 4pm. This is not the case for all currency pairs, which raises the question whether there is a systematic imbalance, which correlates with movements in other markets. 

As a test for such systematic skewing, we have investigated the correlations between the price extrema at 4pm and directional changes in the S\&P 500 equity index. Such correlations could, for instance, exist in pairs containing USD, due to portfolio rebalancing of foreign currency denominated investors in the US market. 
However, the data does not show any significant correlations, which can consistently be attributed to rebalancing in either direction.

\begin{table}[h!]
\begin{center}
  \begin{tabular}{ l | c | c   }
    \toprule
Pair 		& 	$\Delta R(\tilde{t}=\text{16:00})$ maxima [bpts.]	& $\Delta R(\tilde{t}=\text{16:00})$ minima [bpts.] \\
\midrule
AUDUSD	&	$16 \pm 3.5$	&	$-13 	\pm 2.4$		\\
USDCHF	&	$12 \pm 2.5$	&	$-11.3 \pm 2.4$		\\
USDGBP	&	$ 8.3 \pm 1.7$	&	$-8.3 \pm 1.7$		\\
USDJPY	&	$8.7 \pm 2.4$	&	$-9.7 \pm 1.7$		\\
USDSGD	&	$5.3 \pm 1.7$	&	$-5.4 \pm 1.2$		\\
USDMXN	&	$12.3 \pm 2.4$	&	$-10.7 \pm 3.7 $		\\
EURCHF	&	$9.7 \pm 4.2$	&	$-10 \pm 3$		\\
EURJPY	&	$13.3 \pm 4.1$	&	$-14.7 \pm 3.5$		\\
EURUSD	&	$8.3 \pm 2.4$ 	&	$-10.3 \pm 3.5$		\\
EURGBP	&	$9.7 \pm 1.7$	&	$-11 \pm 2.4$		\\
GBPCHF	&	$12 \pm 2.4$	&	$-12.3 \pm 2.4$	\\
\bottomrule
  \end{tabular}
\end{center}
\caption{Average size of extreme interval return movements $\Delta R(\tilde{t} = \text{16:00})$ happening at the centre of the interval 15:40-16:20. The numbers are averages over the movements obtained for the lowest, the highest, and the last price within each 1-min. interval. All three prices types show individual movements of sizes similar to the listed averages.}
\label{tab_ExteremaSize}
\end{table}

\section{Conclusion}

In this article, we reported on the observation of a change in market dynamics around 4pm London time, introduced by the WM/R 4pm fixing rate. 
Our analysis reveals an elevation in volatility, as well as a significant probability for observing extreme price movements within the 1-min. time interval of the WM/R 4pm fixing window. 
These 1-min. movements are not only local extrema but, with a considerable probability, represent the most extreme daily 1-min. movement. 
Both effects happen solely in the 1-min. prior and post 4pm, i.e., they are temporally confined to the WM/R 4pm fixing window. 
This sets the 4pm market dynamics apart from other re-occuring events within the trading day, such as news dissemination or the New York option cut.  

The WM/R 4pm fix thus has its own, special market dynamics, which is caused by 
the compression of a large order flow into a narrow time window and maintained by market participants' behaviour during the fixing period.   

To understand this behaviour and the consequences of our findings, 
we have to consider the usage of the WM/R 4pm fixing rate by market participants, and its usefulness for an investor.

Due to the construction of the fix, risk-averse market makers have the incentive to spread their fixing orders over all 1-sec. intervals within the 1-min. WM/R 4pm fixing window, with preferential execution as close to the end of each 1-sec. time interval as possible. 
By executing closely to each of these 61 time points,
which are the times when the spot rates, going into the WM/R 4pm fix calculation, are measured, 
market makers seek to achieve an average transaction rate over all of these trades that approximates the final WM/R 4pm fixing rate. 
In turn, this trading pattern limits their pricing risks, 
because market makers take on customers' 4pm fixing orders before the WM/R 4pm fixing commences, i.e., they promise a fixed transaction rate without knowing what this price actually is. 
As a consequence of this behaviour, the market is not neccessarily driven by the overall supply and demand of all market participants during the fixing window. 
It is rather dominated by the activities of market-makers, who seek management of the risks they have taken on by agreeing to execute WM/R 4pm fixing trades for investors. 

This raises the question, whether the WM/R 4pm fix is a good benchmark rate to use for an investor.
To this end, let us consider what purposes a fixing benchmark usually serves. 
In principle, it should be the overarching goal of each market participant to achieve best execution when 
trading in the market. 
In reality however, this quest is influenced by the performance assessment mechanisms institutional investors are subject to. 
Investment managers' performance is predominantly measured against benchmark indices. 
So managers have a strong incentive to reduce any tracking error of their portfolio's performance with regard to the respective benchmark indices. 
If a manager's portfolio, and the underlying index, contain FX components, for instance through holding equity denominated in a currency other than the manager's home currency, the index construction usually relies on an FX benchmark rate for currency conversion into the home currency. 
To reduce tracking error, the manager thus also seeks to utilise the same FX benchmark rate, employed by the corresponding index he tracks, for his own currency conversion upon portfolio rebalancing.  
This is particularly the case, if exposure to FX risk is not desired in the construction of the manager's portfolio.  
The by far most popular FX benchmark rate is the WM/R 4pm fix, which explains its importance amongst market participants as well as the large number of orders being traded at the WM/R 4pm fix. 

While usage of a common FX benchmark rate benefits transparency across managers, from our findings, we however conclude, that usage of the WM/R 4pm fixing rate can re-introduce execution risk through the back door.
Due to the higher probability for extreme returns, depending on the investor's trading direction, executing within the fixing window can reduce the investor's overall return, which would not necessarily have been the case for an alternative FX execution.  
Moreover, the increased market volatility around 4pm can also increase the overall volatility of an investors portfolio. 
As a result, WM/R 4pm fix execution can reduce an investor's information ratio, which contradicts the intial intention of best trade execution. 
For these reasons, investors should consider the special market structure in the fixing window, discussed in this article, before blindly relying on the WM/R 4pm fix rate.


\vspace{1.5cm} 

\renewcommand{\baselinestretch}{.50}
\normalsize

\textbf{{\footnotesize DISCLAIMER}} 

\textbf{{\footnotesize Risk Warnings}} 
\newline
{\footnotesize 
All opinion expressed is based on Record’s views as of January 2016 and may have changed since then. The views expressed do not represent financial or legal advice. Record accepts no liability should future events not match these views and strongly recommends you seek your own advice to take account of your specific circumstances. This material is provided for informational purposes only and is not intended to reflect a current or past recommendation, investment advice of any kind, or a solicitation of an offer to buy or sell any securities, Record Currency Management Ltd products or investment services. Any reference to Record products or service is purely incidental and acts as a reference point only for the purposes of this note. The views about the methodology, investment strategy, and its benefits are those held by Record Currency Management Ltd. 

\noindent
All beliefs based on statistical observation must be viewed in the context that past performance is no guide to the future. Changes in rates of exchange between currencies will cause the value of investments to decrease or increase. Before making a decision to invest, you should satisfy yourself that the product is suitable for you by your own assessment, or by seeking professional advice. Your individual facts and circumstances have not been taken into consideration in the production of this document. 
}

\textbf{{\footnotesize Regulated status}} \newline
{\footnotesize 
Record is authorised and regulated by the Financial Conduct Authority in the UK,  registered as an Investment Adviser with the Securities and Exchange Commission in the US, registered as a Commodity Trading Adviser (swaps only) with the US Commodity Futures Trading Commission, is an Exempt International Adviser with the Ontario Securities Commission in Canada, is registered as exempt with the Australian Securities \& Investment Commission, and is approved by the Irish Central Bank to act as promoter and investment manager to Irish authorised collective investment schemes.
}



\renewcommand{\baselinestretch}{1.0}\normalsize

\newpage 

\section{Supplementary Information \label{app_supp_info}}

\subsection{Sizes of interval returns for EURUSD \label{app_size_ext_int_ret}}

Here we provide a brief comparison between the sizes of the extreme interval returns $\tilde{R}_i^j(T_F,\Delta t,d)$ 
and the regular interval returns $R_i^j(t,T_F,\Delta t,d)$. 
Implementing the methodology described in Section \ref{subsec_extr_analysis}, we obtain $N_d$ regular interval returns for each hour $T_F$, as we have $N_d$ trading days. 
The extreme interval returns form a subset thereof, containing $\tilde{N}_d \le N_d$ elements, 
whereby an element is added when the interval return at $T_F$ represents the most extreme interval return throughout all hours within the day $d$. 
Consequently, we end up with a sample set for both, the extreme interval returns and the regular interval returns. 
To compare both variables with one another, prior and post the fixing time $T_F$, for the different price streams (highest, lowest and last price), we calculate
the means of the respective sets.\newline
For the extreme interval returns, these are 
\begin{align*}
\bar{\tilde{R}}^j_i(T_F,\Delta t,d) 	&= \underset{d}{\text{mean}} \left( \left\{| \tilde{R}^j_i(\tilde{T}_F,\Delta t,d) | \right\} \right), \text{and} \\
\delta \bar{\tilde{R}}_i(T_F,\Delta t,d)   &= \underset{d}{\text{mean}} \left( \left\{ \delta \tilde{R}_i(\tilde{T}_F,\Delta t,d) \right\} \right), 
\end{align*} 
while for the regular interval returns the expressions read 
\begin{align*}
\bar{R}^j_i(t,T_F,\Delta t,d) 	&= \underset{d}{\text{mean}} \left( \left\{ | R^{j}_i(t,T_F,\Delta t,d) | \right\} \right), \text{and} \\
\delta \bar{R}_i(t,T_F,\Delta t,d) &= \underset{d}{\text{mean}} \left( \left\{ | \delta R_i(t,T_F,\Delta t,d) | \right\} \right).
\end{align*} 
\noindent
Notably, we take the absolute value of the interval returns, as these can be positive or negative. 
In other words, there are situations when the spot rate just keeps 
increasing or decreasing throughout the time period $\left[T_F - \Delta t, T_F + \Delta t\right]$, as well as occasions when the sport rates appreciates in 
$\left[T_F - \Delta t, T_F\right]$ just to fall again in $\left[T_F, T_F + \Delta t\right]$, and vice versa. 
These four scenarios occur with approximately equal frequency within the datasets.
Moreover, the return sizes clearly depend on the interval size $\Delta t$ and the currency pair. 
As we are mainly concerned with the dynamics in the WM/R 4pm fixing window, we quote the samples means considering a
$\Delta t= 1\text{-min.}$ interval around $T_F = \text{16:00}$ for the pair EURUSD, 
which is our example of choice to discuss our results in Section \ref{sec_ext_ret_distr_observations}.
We obtain the values stated in table \ref{tab_app_size_ext_ret_eurusd}.
\begin{table}[h!]
\centering
\begin{tabular}{l|c|c|c|c|c|c}
\toprule
Interval return					& Int $i$	&Type $j$	& High 	& Low 	& Last 	& High - low 	\\
\midrule
$\bar{\tilde{R}}_i^j(T_F,\Delta t)$	& int1		& max		& $4.8$	& $-$ 		& $5.0$ 	& $-$	\\
$\bar{\tilde{R}}(T_F,\Delta t)$		& int2		& max		& $4.6$	& $-$ 		& $4.6$ 	& $-$	\\	
$\bar{R}(t,T_F,\Delta t)$			& int1		& max		& $1.3$ 	& $-$ 		& $1.3$ 	& $-$	\\
$\bar{R}(t,T_F,\Delta t)$			& int2		& max		& $0.6$	& $-$ 		& $0.9$ 	& $-$	\\
\midrule
$\bar{\tilde{R}}(T_F,\Delta t)$		& int1		& min		& $-$		& $5.0$ 	& $4.9$ 	& $-$	\\
$\bar{\tilde{R}}(T_F,\Delta t)$		& int2		& min		& $-$		& $4.2$ 	& $4.2$ 	& $-$	\\
$\bar{R}(t,T_F,\Delta t)$			& int1		& min		& $-$		& $1.3$ 	& $1.2$ 	& $-$	\\
$\bar{R}(t,T_F,\Delta t)$			& int2		& min		& $-$		& $0.6$ 	& $0.8$ 	& $-$	\\
\midrule
$\delta \bar{\tilde{R}}(T_F,\Delta t)$	& int1		& diff		& $-$		& $-$ 		& $-$ 		& $11.6$	\\
$\delta\bar{\tilde{R}}(T_F,\Delta t)$	& int2		& diff		& $-$		& $-$ 		& $-$ 		& $12.4$	\\
$\delta \bar{R}(t,T_F,\Delta t)$		& int1		& diff		& $-$		& $-$ 		& $-$ 		& $7.5$	\\
$\delta \bar{R}(t,T_F,\Delta t)$		& int2		& diff		& $-$		& $-$ 		& $-$ 		& $7.3$	\\
\bottomrule
\end{tabular}
\caption{Average sizes of the extreme interval returns $\bar{\tilde{R}}_i^j(T_F,\Delta t,d)$ and the regular interval returns $\bar{R}_i^j(t,T_F,\Delta t,d)$, as well as return differences $\delta \bar{\tilde{R}}_i(T_F,\Delta t,d)$ and $\delta \bar{R}_i(t,T_F,\Delta t,d)$ for EURUSD.
We consider a time window of $\Delta t = 1\text{-min.}$ around $T_F=$16:00, representing the WM/R 4pm fixing time. All numbers have units in basis points (1/100 of a percent).}
\label{tab_app_size_ext_ret_eurusd}
\end{table}

\subsection{Extreme period return probability distributions for all 12 currency pairs\label{app_prob_ext_ret}}

In Figure \ref{fig_3} of Section \ref{sec_ext_ret_distr_observations}, we have shown the probability distribution $P(T_F,\Delta t)$ of the most extreme period returns
for the currency pair EURUSD. 
To complete this information with the data for the other 11 studied currency pairs, we present two cross sections of the 3-dimensional probability surfaces $P(T_F,\Delta t)$.
 
\begin{figure}[h!]
\centering
\includegraphics[width=0.98\textwidth]{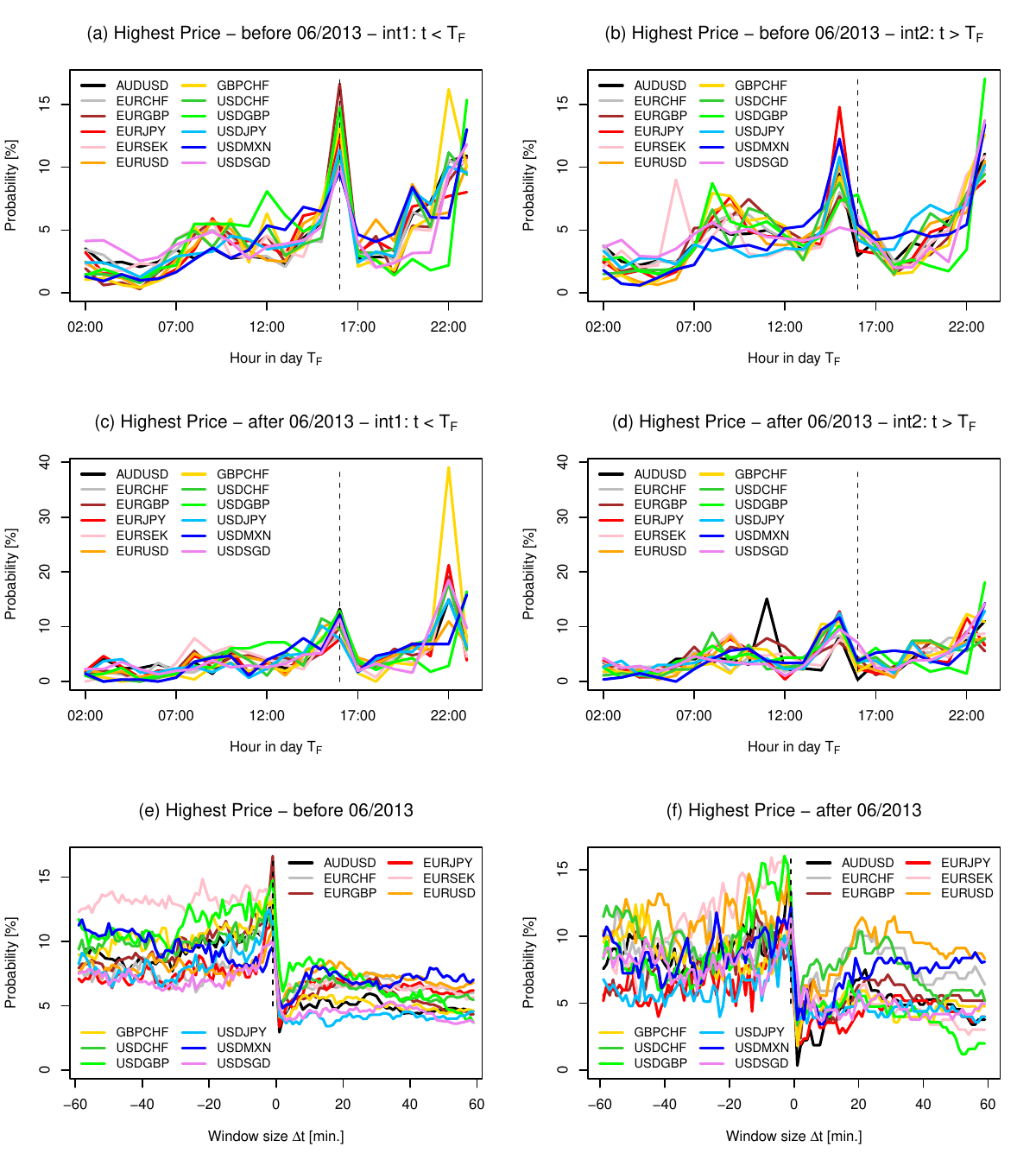}
\caption{
Cross-sections of the probability distribution $P(T_F,\Delta t)$ for the highest price in each 1-min. timestep. 
(a)-(d) show $P(T_F,\Delta t = 1-min.)$ for the minimal interval size $\Delta t$ against the hours in the day $T_F$. 
Times $t$ in int1 before the fixing hour, i.e. $t \in \left[ T_F-\Delta t,T_F\right]$ are displayed in (a) and (c), with times thereafter, i.e. 
$t \in \left[ T_F, T_F+\Delta t\right]$, shown in (b) and (d). 
Plots (e) and (f) present $P(T_F = \text{16:00},\Delta t)$ as a function of window size $\Delta t$ at the time of the WM/R 4pm fix. The intervals before and after the fix, int1 and int2, are denoted by negative and positive $\Delta t$-values, respectively. 
Plots (a), (b), (e) contain data before 06/2013, whereas data from 06/2013 onwards is shown in the plots (c), (d), (f).}
\label{supp_fig_1}
\end{figure}

\begin{figure}[h!]
\centering
\includegraphics[width=0.98\textwidth]{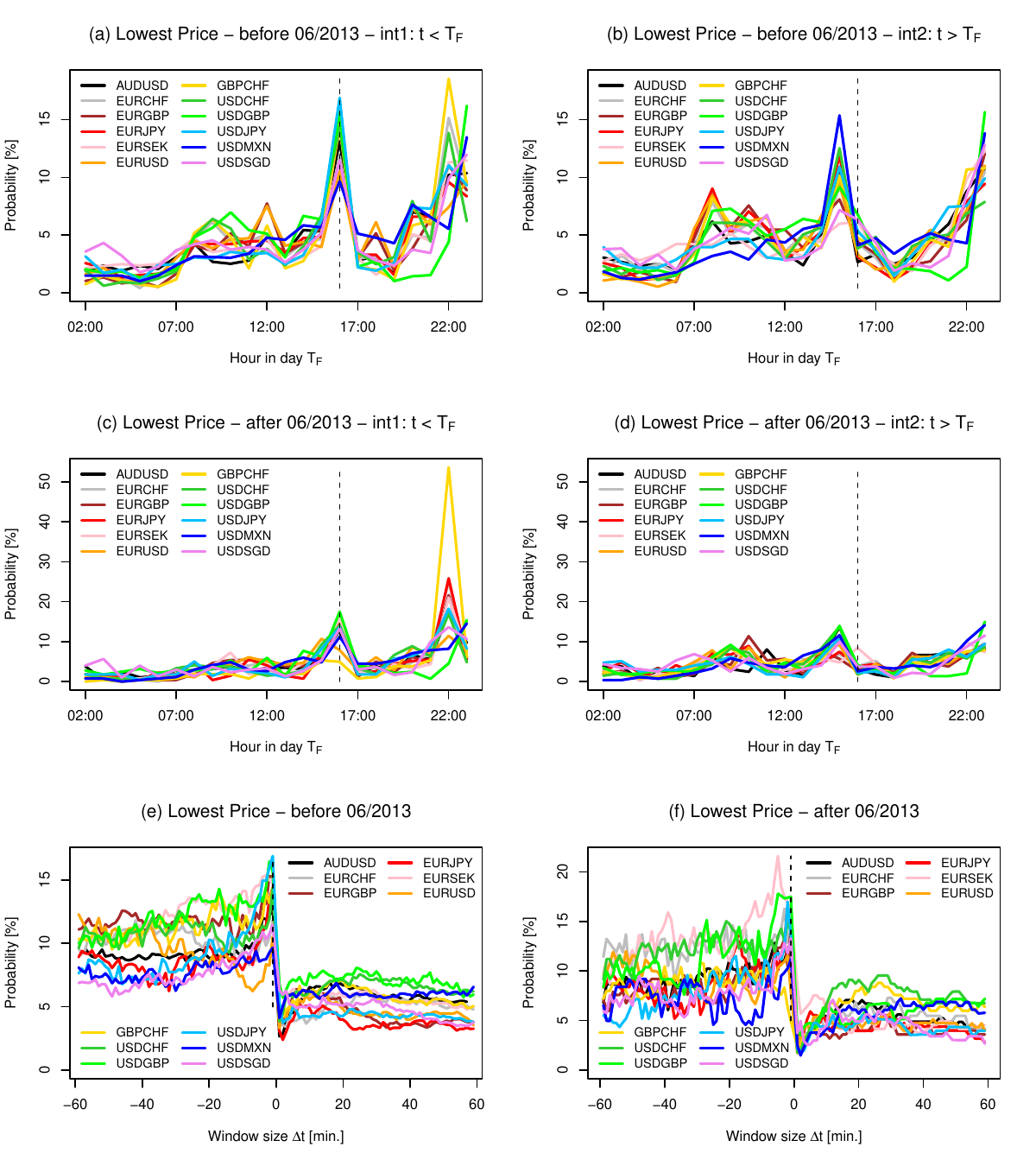}
\caption{
Cross-sections of the probability distribution $P(T_F,\Delta t)$ for the lowest price in each 1-min. timestep. 
(a)-(d) show $P(T_F,\Delta t = 1-min.)$ for the minimal interval size $\Delta t$ against the hours in the day $T_F$. 
Times $t$ in int1 before the fixing hour, i.e. $t \in \left[ T_F-\Delta t,T_F\right]$ are displayed in (a) and (c), with times thereafter, i.e. 
$t \in \left[ T_F,T_F+\Delta t\right]$, shown in (b) and (d). 
Plots (e) and (f) present $P(T_F = \text{16:00},\Delta t)$ as a function of window size $\Delta t$ at the time of the WM/R 4pm fix. The intervals before and after the fix, int1 and int2, are denoted by negative and positive $\Delta t$-values, respectively. 
Plots (a), (b), (e) contain data before 06/2013, whereas data from 06/2013 onwards is shown in the plots (c), (d), (f).
}
\label{supp_fig_2}
\end{figure}

\begin{figure}[h!]
\centering
\includegraphics[width=0.98\textwidth]{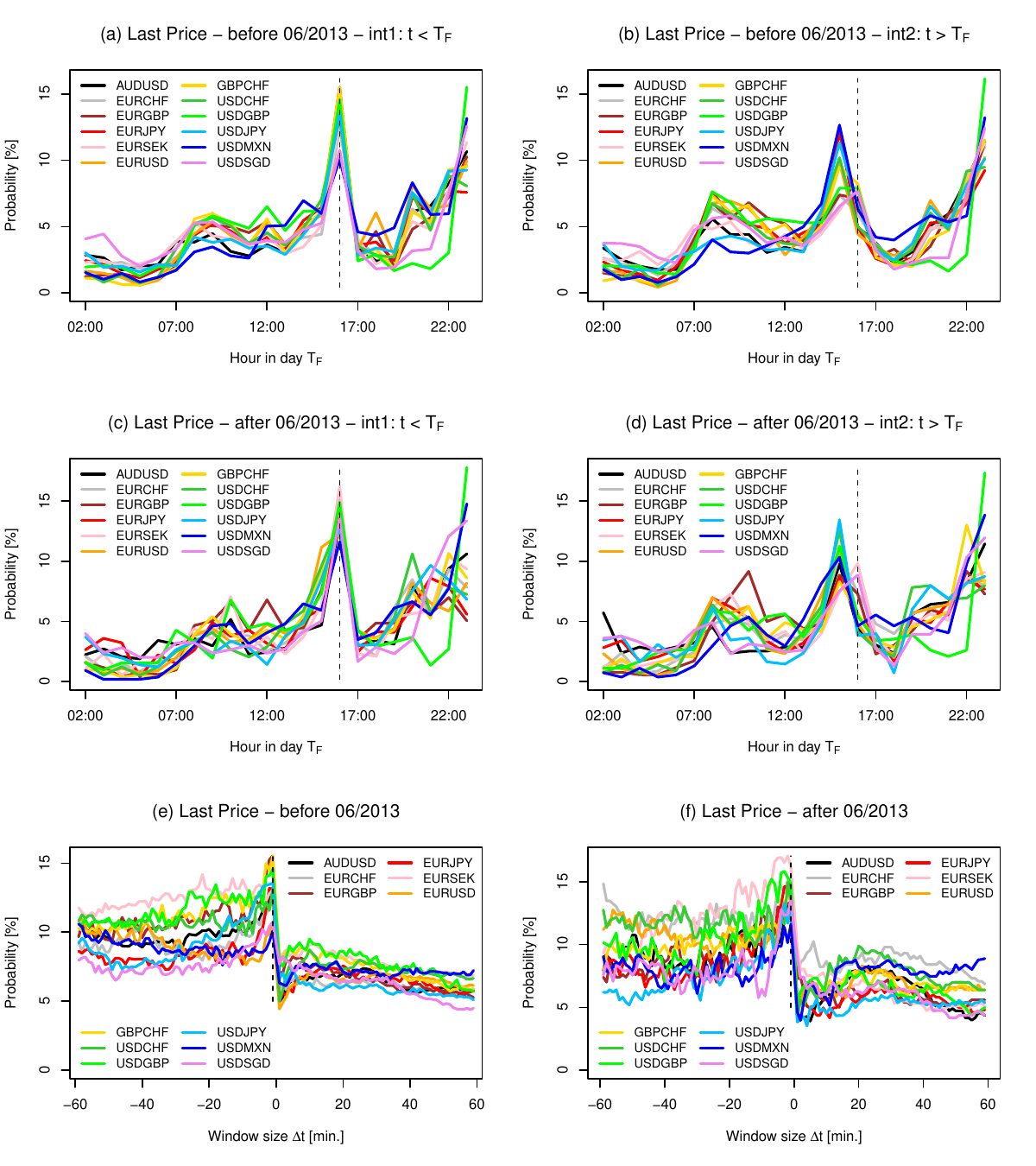}
\caption{
Cross-sections of the probability distribution $P(T_F,\Delta t)$ for the last price in each 1-min. timestep. 
(a)-(d) show $P(T_F,\Delta t = 1-min.)$ for the minimal interval size $\Delta t$ against the hours in the day $T_F$. 
Times $t$ in int1 before the fixing hour, i.e. $t \in \left[ T_F-\Delta t,T_F\right]$ are displayed in (a) and (c), with times thereafter, i.e. 
$t \in \left[ T_F,T_F+\Delta t\right]$, shown in (b) and (d). 
Plots (e) and (f) present $P(T_F = \text{16:00},\Delta t)$ as a function of window size $\Delta t$ at the time of the WM/R 4pm fix. The intervals before and after the fix, int1 and int2, are denoted by negative and positive $\Delta t$-values, respectively. 
Plots (a), (b), (e) contain data before 06/2013, whereas data from 06/2013 onwards is shown in the plots (c), (d), (f).
}
\label{supp_fig_0}
\end{figure}

\begin{figure}[h!]
\centering
\includegraphics[width=0.95\textwidth]{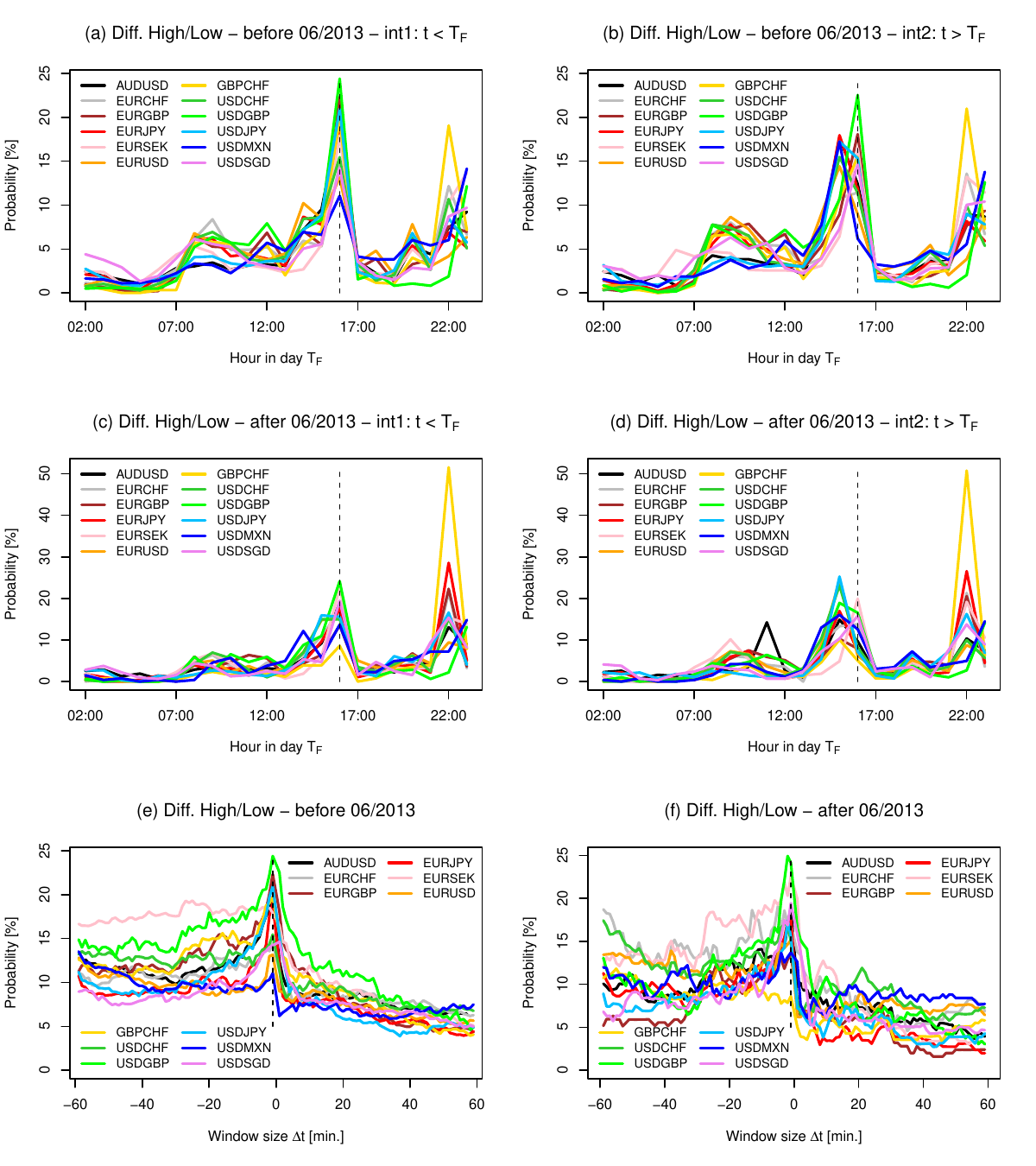}
\caption{
Cross-sections of the probability distribution $P(T_F,\Delta t)$ for the difference between the highest and the lowest price in each 1-min. timestep. 
(a)-(d) show $P(T_F,\Delta t = 1-min.)$ for the minimal interval size $\Delta t$ against the hours in the day $T_F$. 
Times $t$ in int1 before the fixing hour, i.e. $t \in \left[ T_F-\Delta t,T_F\right]$ are displayed in (a) and (c), with times thereafter, i.e. 
$t \in \left[ T_F,T_F+\Delta t\right]$, shown in (b) and (d). 
Plots (e) and (f) present $P(T_F = \text{16:00},\Delta t)$ as a function of window size $\Delta t$ at the time of the WM/R 4pm fix. The intervals before and after the fix, int1 and int2, are denoted by negative and positive $\Delta t$-values, respectively. 
Plots (a), (b), (e) contain data before 06/2013, whereas data from 06/2013 onwards is shown in the plots (c), (d), (f).
}
\label{supp_fig_3}
\end{figure}

For the first cross-section, we fix the interval to the shortest size of $\Delta t= 1\text{-min}$. 
Here, the probability spikes, induced by the market dynamics of the WM/R fixing window prior to $T_F = \text{16:00}$, show their greatest influence. 
We consequently plot ${P_i(T_F) := P_i(T_F,\Delta t = 1\text{-min.})}$ as a function of the time $T_F$ in the trading day, 
with the results for all currency pairs shown in Figures \ref{supp_fig_1}-\ref{supp_fig_3} for the highest prices, the lowest prices, the last price, and the difference between the highest and lowest prices, respectively.  
Similar to Figure \ref{fig_3}, this is done separately for the intervals $i \in \left\{ \text{int1},\text{int2} \right\}$ before and after the respective fixing time $T_F$, with data for int1 shown in the left columns and data for int2 shown in the right columns of Figures  \ref{supp_fig_1}-\ref{supp_fig_3} subplots (a)-(d). 
The period prior to June 2013 is displayed in subplots (a) and (b), whereas subplots (c) and (d) contain data for the time thereafter. 

With the second cross-section, we demonstrate the localisation of the probability of extreme interval returns at the WM/R 4pm fix. 
Here, we keep the fixing time $T_F$ constant and display the scaling of 
${P_i(\Delta t) = P_i(T_F = \text{16:00},\Delta t)}$ as a function of the interval size $\Delta t$. 
Both intervals $i \in \left\{ \text{int1},\text{int2} \right\}$ are combined into a single plot, whereby we denote the interval prior to 4pm (int1) using negative values for the interval size $\Delta t$. 
Data prior to June 2013 is presented in the subplots (e) of Figures \ref{supp_fig_1}-\ref{supp_fig_3}, while the time thereafter is shown in subplots (f).

\subsection{Distribution of extreme spot movements in 40-min. intervals for all 12 currency pairs\label{app_hist_ext_ret_at_4pm}}

\begin{figure}[h!]
\centering
\includegraphics[width=1\textwidth]{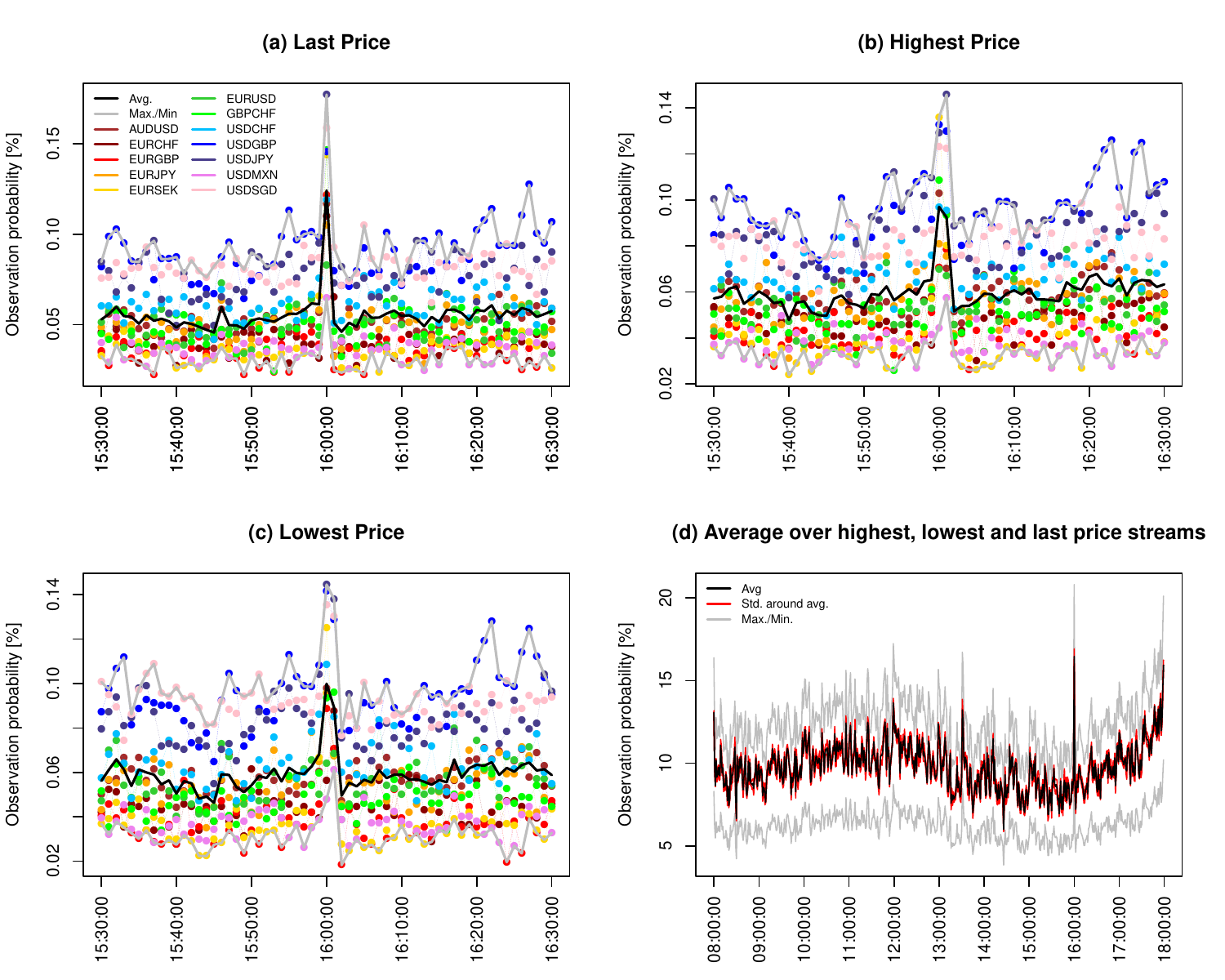}
\caption{
Probability $P(\tilde{t})$ for extreme returns, centred in a 40-min. interval around a fixing time $\tilde{t} = T_F$, with $T_F$ moved through every minute within a trading day. 
(a)-(c) show the individual price streams of highest, lowest and last price within the 1-min. time intervals of our datasets, focussing on the time around the WM/R 4pm fix. Coloured datapoints represent the individual currency pairs, while the black line denotes average probability over all pairs, and the grey lines mark the overall minimum and maximum probabilities. The values $P(\tilde{t})$ are calculated using the sum of extreme positive and negative returns at $\tilde{t}$. 
(d) displays the average probability (black), its $\pm 1 \sigma$ confidence bound (red) alongside the maximum and minimum probability over all currency pairs (grey), for the trading day. 
}
\label{supp_fig_5}
\end{figure}

Using the methodology outlined in Section \ref{subsec_corr_method}, we present the data for the numbers $N^j(\tilde{t})$ of maximal/minimal period returns $R^j(\tilde{t})$, with $j\in \left\{\text{max}, \text{min}\right\}$, observed at time $\tilde{t}$, for spot rates within an interval of size $\Delta t =  20\text{-min.}$ to either side of $\tilde{t}$, centred at $\tilde{t}$.
Similar to the USDJPY example of Figure \ref{fig_4} (b), we plot the histograms for the number of times these extreme period returns have occured at each minute within a trading day. 
The data for all 12 currency pairs, subject to our study, is shown in Figure \ref{supp_fig_5} for the three price streams of highest, lowest and last price in each 1-min. time step.

The numbers of maxima and minima in period returns observed at the WM/R 4pm fixing time, 
$N^j_\text{4pm} = N^j(\tilde{t} = \text{16:00})$, 
as well as the average number of such extrema 
$N^j_{\neq \text{4pm}} = N^j(\tilde{t} \neq \text{16:00})$  
occuring at any other time of the day, are summarised in Tables \ref{tab_supp_1} - \ref{tab_supp_3} for the 
last, the highest, and the lowest price. 

\begin{table}[h!]
\begin{center}
  \begin{tabular}{ l | c | c | c | c | c | c  }
    \toprule
 	&	\multicolumn{3}{c|}{Time $\tilde{t} = \text{16:00}$}		& 	 \multicolumn{3}{c}{Time $\tilde{t} \neq \text{16:00}$} \\
Pair 	& 	$N(\tilde{t})$ &  $N_d$ & $P(\tilde{t})$ [\%] 	& $N(\tilde{t})$ &  $N_d$ 	& $P(\tilde{t})$ [\%]\\
\midrule
AUDUSD	 & $161$	& $1381$	& $11.7$	& $81 \pm13 $	& $1383 \pm 1$		& $5.9 \pm 0.9$	\\
EURCHF	 & $84$	& $766$	& $11.0$	& $32 \pm 6$	& $766.00 \pm 0.05 $	& $4.2 \pm 0.7$	\\
EURGBP	 & $93$	& $766$	& $12.1$	& $28 \pm 6$	& $765.9 \pm 0.3$		& $3.7 \pm 0.8$	\\
EURJPY	 & $78$	& $743$	& $10.5$	& $39 \pm 8$	& $743$			& $5.3 \pm 1.1$	\\
EURSEK	 & $199$	& $1384$	& $14.4$	& $53 \pm 10$	& $1384 \pm 1$		& $3.8 \pm 0.7$	\\
EURUSD	 & $58$	& $702$	& $8.3$	& $38 \pm 8$	& $702.0 \pm 0.3$		& $5.3 \pm 1.1$	\\
GBPCHF	 & $108$	& $738$	& $14.6$	& $37 \pm 7$	& $741 \pm 1$		& $5.0 \pm 1.0$	\\
USDCHF	 & $91$	& $766$	& $11.9$	& $51 \pm 9$	& $766.0 \pm 0.1$		& $6.6 \pm 1.2$	\\
USDGBP 	 & $113$	& $777$	& $14.5$	& $84 \pm 10$	& $774 \pm 13$		& $10.5 \pm 2.1$	\\
USDJPY	 & $136$	& $768$	& $17.7$	& $70 \pm 13$	& $768.00 \pm 0.04$	& $9.2 \pm 1.7$	\\
USDMXN 	 & $89$	& $1376$	& $6.5$	& $54 \pm 10$	& $1312 \pm 79$		& $4.1 \pm 0.8$	\\
USDSGD	 & $219$	& $1382$	& $15.8$	& $125 \pm 22$	& $1376 \pm 10$		& $9.1 \pm 1.6$	\\
\bottomrule
  \end{tabular}
\end{center}
\caption{
Number $N(\tilde{t})$ and probability $P(\tilde{t})$ of extreme returns in a $\Delta t = 20\text{-min.}$ interval for the last price in each 1-min. time interval during a trading day.$N_d$ is the total number of trading days in the sample. The time $\tilde{t} = \text{16:00}$ is the interval centred at the WM/R 4pm fixing time, whereas $\tilde{t} \neq \text{16:00}$ denotes the average over all other minutes within a day. 
Both numbers are the sum of positive and negative interval returns. 
}
\label{tab_supp_1}
\end{table}

\begin{table}[h!]
\begin{center}
  \begin{tabular}{ l | c | c | c | c | c | c  }
    \toprule
 	&	\multicolumn{3}{c|}{Time $\tilde{t} = \text{16:00}$}		& 	 \multicolumn{3}{c}{Time $\tilde{t} \neq \text{16:00}$} \\
Pair 	& 	$N(\tilde{t})$ &  $N_d$ & $P(\tilde{t})$ [\%] 	& $N(\tilde{t})$ &  $N_d$ 	& $P(\tilde{t})$ [\%]\\
\midrule
AUDUSD	 & $142$	& $1381$	& $10.3$	& $88 \pm 13$	& $1383 \pm 1$		& $ 6.4 \pm 0.94$	\\
EURCHF	 & $54$	& $766$	& $7.1$	& $41 \pm 6$	& $766.00 \pm 0.05$	& $5.3 \pm 0.9$	\\
EURGBP	 & $54$	& $766$	& $7.0$	& $35 \pm 7$	& $765.9 \pm 0.1$		& $4.6 \pm 0.9$	\\
EURJPY	 & $60$	& $743$	& $8.1$	& $43 \pm 8$	& $743$			& $5.8 \pm 1.1$	\\
EURSEK	 & $188$	& $1384$	& $13.6$	& $54 \pm 10$	& $1384 \pm 1$		& $3.9 \pm 0.7$	\\
EURUSD	 & $49$	& $702$	& $7.0$	& $41 \pm 8$	& $702.0 \pm 0.3$		& $5.8 \pm 1.1$	\\
GBPCHF	 & $80$	& $738$	& $10.8$	& $39 \pm 8$	& $741 \pm 1$		& $5.3 \pm 1.0$	\\
USDCHF	 & $74$	& $766$	& $9.7$	& $56 \pm 9$	& $766.0 \pm 0.1$		& $7.3 \pm 1.2$	\\
USDGBP	 & $103$	& $777$	& $13.3$	& $85 \pm 17$	& $774 \pm 13$		& $11.0 \pm 2.2$	\\
USDJPY	 & $99$	& $768$	& $12.9$	& $75 \pm 12$	& $768.00 \pm 0.04$	& $9.8 \pm 1.5$	\\
USDMXN	 & $61$	& $1376$	& $4.4$	& $53 \pm 9$	& $1312 \pm 79$		& $4.1 \pm 0.7$	\\
USDSGD	 & $170$	& $1382$	& $12.3$	& $126 \pm 21$	& $1376 \pm 10$		& $9.2 \pm 1.5$	\\
\bottomrule
  \end{tabular}
\end{center}
\caption{Number $N(\tilde{t})$ and probability $P(\tilde{t})$ of extreme returns in a $\Delta t = 20\text{-min.}$ interval for the last price in each 1-min. time interval during a trading day. $N_d$ is the total number of trading days in the sample. The time $\tilde{t} = \text{16:00}$ is the interval centred at the WM/R 4pm fixing time, whereas $\tilde{t} \neq \text{16:00}$ denotes the average over all other minutes within a day. 
Both numbers are the sum of positive and negative interval returns.}
\label{tab_supp_2}
\end{table}

\begin{table}[h!]
\begin{center}
  \begin{tabular}{ l | c | c | c | c | c | c  }
    \toprule
 	&	\multicolumn{3}{c|}{Time $\tilde{t} = \text{16:00}$}		& 	 \multicolumn{3}{c}{Time $\tilde{t} \neq \text{16:00}$} \\
Pair 	& 	$N(\tilde{t})$ &  $N_d$ & $P(\tilde{t})$ [\%] 	& $N(\tilde{t})$ &  $N_d$ 	& $P(\tilde{t})$ [\%]\\
\midrule
AUDUSD	 & $135$	& $1381$	& $9.8$	& $90 \pm 14 $	& $1383 \pm 1$		& $6.5 \pm 1.0$	\\
EURCHF 	 & $ 43$	& $766$	& $5.6$	& $36 \pm 6$	& $766.00 \pm 0.05$	& $4.7 \pm 0.8$	\\
EURGBP	 & $68$	& $766$	& $8.9$	& $29 \pm 6$	& $765.9 \pm 0.3$		& $3.8 \pm 0.8$	\\
EURJPY	 & $71$	& $743$	& $9.6$	& $42 \pm 8$	& $743$			& $5.7 \pm 1.1$	\\
EURSEK	 & $173$	& $1384$	& $12.5$	& $53 \pm9 $	& $1384 \pm 1$		& $3.8 \pm 0.6$	\\
EURUSD  	 & $45$	& $702$	& $6.4$	& $41 \pm 8$	& $702.0 \pm 0.3$		& $5.9 \pm 1.1$	\\
GBPCHF	 & $69$	& $738$	& $9.3$	& $39 \pm 7$	& $741 \pm 1$		& $5.2 \pm 1.0$	\\
USDCHF	 & $83$	& $766$	& $10.8$	& $54 \pm 9$	& $766.0 \pm 0.1$		& $7.1 \pm 1.2$	\\
USDGBP	 & $110$	& $777$	& $14.2$	& $85 \pm 16$	& $774 \pm 13$		& $11.0 \pm 2.1$	\\
USDJPY	 & $111$	& $768$	& $14.5$	& $74 \pm 13$	& $768.00 \pm 0.04$	& $9.7 \pm 1.6$	\\
USDMXN	 & $66$	& $1376$	& $4.8$	& $54 \pm 9$	& $1312 \pm 79$		& $4.1 \pm 0.7$	\\
USDSGD	 & $187$	& $1382$	& $13.5$	& $138 \pm 23$	& $1376 \pm 10$		& $10.0 \pm 1.7$	\\
\bottomrule
  \end{tabular}
\end{center}
\caption{
Number $N(\tilde{t})$ and probability $P(\tilde{t})$ of extreme returns in a $\Delta t= 20\text{-min.}$ interval for the last price in each 1-min. time interval during a trading day. $N_d$ is the total number of trading days in the sample. The time $\tilde{t} = \text{16:00}$ is the interval centred at the WM/R 4pm fixing time, whereas $\tilde{t} \neq \text{16:00}$ denotes the average over all other minutes within a day. 
Both numbers are the sum of positive and negative interval returns.
}
\label{tab_supp_3}
\end{table}

The numbers for each price stream represent the sum of positive and negative return extrema, i.e.,  
$N(\tilde{t}) = N^\text{max}(\tilde{t}) + N^\text{min}(\tilde{t})$. 
$N_d$ is the total number of days in each dataset for which the number of extreme interval returns is counted.
Since the numbers for $N^j(\tilde{t} \neq \text{16:00})$ are mean values over all minutes $\tilde{t} \neq \text{16:00}$, they carry an error, following from the standard deviation of the observed number of maxima at all such minutes $\tilde{t}$ within a day. 
Conversely, the number of 
$N(\tilde{t}=\text{16:00})$ is an exact number, because it denotes the counts for a single point in time ($\tilde{t} = \text{16:00}$). 
Similar to the results in Table \ref{tab_ExteremaNumber}, we calculate the probability to observe extrema in the interval returns as $P(\tilde{t}) = \frac{N(\tilde{t})}{N_d}$. 
Note furthermore that we are not restricting the datasets for each day to contain entries for all minutes within the day as we did for the analysis presented in Section \ref{sec_extr}. 
Here, we keep the time interval $\Delta t = 20\text{-min.}$ fixed and do not directly compare the number of observed extrema $N(\tilde{t})$ for different times $\tilde{t}$ within the day, as we did in Section \ref{sec_extr}. 
Instead, we compare the probabilties $P(\tilde{t})$, which are normalised to the number of daily datasets available at $\tilde{t}$. These are thus independent of a variation in $N_d$ between times $\tilde{t}=\text{16:00}$ and $\tilde{t} \neq \text{16:00}$. 
Consequently, we quote two different numbers for $N_d$, $N_d(\tilde{t} = \text{16:00})$, and $N_d(\tilde{t} \neq \text{16:00})$ in 
Table \ref{tab_ExteremaNumber} of the main text and Tables \ref{tab_supp_1} - \ref{tab_supp_3}.

\begin{table}[h!]
\begin{center}
  \begin{tabular}{ l | c | c | c | c | c | c  }
    \toprule
Pair & 	\multicolumn{2}{c|}{Last Price} & \multicolumn{2}{c|}{Highest Price} & \multicolumn{2}{c}{Lowest Price} \\
&	$\Delta R^\text{max}$ &  $\Delta R^\text{min}$ & $\Delta R^\text{max}$ & 
$\Delta R^\text{min}$ &  $\Delta R^\text{max}$	& 	$\Delta R^\text{min}$  \\
\midrule
AUDUSD	 & $17 \pm 2$	& $-14 \pm 1$	& $16 \pm 2$	& $-12 \pm 2$	& $14 \pm 2$	& $-13 \pm 1$	\\
EURCHF	 & $8 \pm 1$	 	& $-10 \pm 1$	& $9 \pm 1$		& $-9 \pm 2$		& $12 \pm 4$	& $-11 \pm 2$	\\
EURGBP	 & $11 \pm 1$	& $-10 \pm 1$	& $10 \pm 1$	& $-11 \pm 2$	& $8 \pm 1$		& $-12 \pm 1$	\\
EURJPY	 & $14 \pm 2$	& $-16 \pm 2$ 	& $12 \pm 2$	& $-13 \pm 2$	& $14 \pm 3$	& $-15 \pm 2$	\\
EURUSD	 & $8 \pm 1$		& $-11 \pm 2$	& $7 \pm 1$		& $-10 \pm 2$	& $9 \pm 2$		& $-10 \pm 2$	\\
GBPCHF	 & $12 \pm 1$	& $-13 \pm 1$	& $12 \pm 1$	& $-11 \pm 2$	& $12 \pm 2$	& $-13 \pm 1$ 	\\
USDCHF	 & $13 \pm 1$	& $-12 \pm 2$	& $12 \pm 1$	& $-11 \pm 2$	& $11 \pm 2$	& $-11 \pm 1$	\\
USDGBP	 & $9 \pm 1$  	& $-10 \pm 1$	& $8 \pm 1$		& $-7 \pm 1$		& $8 \pm 1$		& $-8 \pm 1$ 	\\
USDJPY	 & $9 \pm 1$  	& $-10 \pm 1$	& $10 \pm 1$	& $-10 \pm 1$ 	& $8 \pm 2$		& $-9 \pm 1$		\\
USDMXN	 & $11 \pm 1$ 	& $-10 \pm 1$ 	& $13 \pm 1$	& $-12 \pm 3$	& $13 \pm 2$	& $-10 \pm 2$	\\
USDSGD	 & $6 \pm 1$	  	& $-6 \pm 1$		& $6 \pm 1$		& $-5 \pm 1$		& $4 \pm 1$		& $-6 \pm 1$		\\
\bottomrule
  \end{tabular}
\end{center}
\caption{Sizes of extreme interval return movements $\Delta R^j(\tilde{t} = \text{16:00})$ for the last, the highest and the lowest price quotes in each 1-min. interval for all 12 currency pairs. All values for $\Delta R^j$ are in units of bpts.}
\label{tab_supp_4}
\end{table}

\subsection{Size of extreme spot movements in 40-min. intervals for all 12 currency pairs\label{app_size_ext_ret_at_4pm}} 
The sizes of the extreme period returns $\Delta R^j(\tilde{t} = \text{16:00})$, quoted in Table \ref{tab_ExteremaSize} of Section \ref{subsec_size_results} in the main text, are the averages over all three price streams of highest, lowest, and last price per 1-min. time step. 
The underlying data for each separate price stream is stated in Table \ref{tab_supp_4}. 
The errors on the quoted numbers in Table \ref{tab_supp_4} derive from the standard deviation of all observed extreme interval return sizes $\Delta R^j(\tilde{t} = \text{16:00})$.

\end{document}